\documentclass[10pt]{article}
\usepackage[top=2cm,left=2cm,right=2cm]{geometry}
\usepackage[numbers,square]{natbib}

\usepackage[english]{babel}
\usepackage[latin1]{inputenc}
\usepackage{hyperref}
\usepackage{amssymb}
\usepackage{amsfonts}
\usepackage{amsmath}
\usepackage{fancyhdr}
\usepackage{epsfig}
\usepackage{graphics}
\usepackage{subfigure}
\usepackage{rotating}
\usepackage{lineno}
\usepackage{pstricks}
\usepackage{color}

\let\a = \alpha    \let\b =\beta            \let\z =\zeta    
                  
                    \let\r = \rho    \let\s = \sigma

\renewcommand{\Im}{\mathrm{Im}}
\newcommand{\bq}{\bar{q}}

\newcommand{\muu}{\mu_{1}}
\newcommand{\nuu}{\nu_{1}}
\newcommand{\ku}{\vec{k_1}}
\newcommand{\xu}{x_{1}}

\newcommand{\mud}{\mu_{2}}
\newcommand{\nud}{\nu_{2}}
\newcommand{\kd}{\vec{k_2}}
\newcommand{\xd}{x_{2}}

\newcommand{\mut}{\mu_{3}}
\newcommand{\nut}{\nu_{3}}
\newcommand{\kt}{\vec{k_3}}
\newcommand{\xt}{x_{3}}

\newcommand{\muq}{\mu_{4}}
\newcommand{\nuq}{\nu_{4}}
\newcommand{\kq}{\vec{k_4}}
\newcommand{\xq}{x_{4}}

\newcommand{\beq}{\begin{equation}}
\newcommand{\eeq}{\end{equation}}
\newcommand{\beqa}{\begin{eqnarray}}
\newcommand{\eeqa}{\end{eqnarray}}

\newcommand{\nn}{\nonumber}

\newcommand{\pd}{\partial}

\newcommand{\bea}{\begin{eqnarray}}
\newcommand{\eea}{\end{eqnarray}}

\begin{document}
\begin{center}
\vspace{4.cm}
{\bf \large 
Three and Four Point Functions of Stress Energy Tensors in $D=3$ \\
for the Analysis of Cosmological Non-Gaussianities}

\vspace{1cm}

{\bf Claudio Corian\`{o}, Luigi Delle Rose and Mirko Serino}

\vspace{1cm}

{  Dipartimento di Matematica e Fisica {\em "Ennio De Giorgi"},\\
Universit\`{a} del Salento \\ and \\ INFN Lecce, Via Arnesano, 73100 Lecce, Italy\\}
\vspace{0.5cm}

\begin{abstract}
We compute the correlation functions of 3 and 4 stress energy tensors $(T)$ in $D=3$ in free field theories of scalars, abelian gauge fields, and fermions, which are relevant in the analysis of cosmological non-gaussianities.
These correlators appear in the holographic expressions of the scalar and tensor perturbations 
derived for holographic cosmological models. The result is simply adapted to describe the leading contributions in the gauge coupling to the same correlators 
also for a non abelian SU(N) gauge theory.  In the case of the bispectrum, our results are mapped and shown to be in full agreement with the corresponding expressions given in a recent holographic study by Bzowski, McFadden and Skenderis. In the 4-T case we present the completely traced amplitude plus all the contact terms. These are expected to appear in a fourth order extension of the holographic  formulas for the 4-point functions of scalar metric perturbations.

\end{abstract}
\end{center}
 \newpage
\section{Introduction} 
Conformal field theories in $D>2$ are significantly less known compared to their $D=2$ counterparts, where exact results stemming 
from the presence of an underlying infinite dimensional symmetry have allowed to proceed with their classification. In fact, as one 
moves to higher spacetime dimensions, 
the finite dimensional character of the conformal symmetry allows to fix, modulo some overall constants, only the structure of 2- and 
3-point functions \cite{Osborn:1993cr, Erdmenger:1996yc}. 
In 4-D, for instance, free field theory realizations of these specific correlators allow the identification of their explicit 
expressions, performing a direct comparison with their general form, which is predicted by the symmetry \cite{Coriano:2012wp}. 
Among these correlators, a special role is taken by those involving insertions of the energy momentum tensor (EMT), which can be 
significant in the context of several phenomenological applications. For instance, in $D=4$,  correlators involving insertions of the 
EMT describe the interaction of a given theory with gravity around the flat spacetime limit. Their study is quite involved due to the 
appearance of a trace anomaly \cite{Giannotti:2008cv, Armillis:2009pq, Coriano:2011ti, Coriano:2011zk}. They are part of the 
anomalous effective action of gravitons at higher order, but they also find application in the description of dilaton interactions 
and of the Higgs-dilaton mixing at the LHC \cite{Coriano:2012nm, Coriano:2012dg}.

In 3 dimensions their computation simplifies considerably, due to the absence of anomalies, 
but it remains quite significant, especially in the context of the ADS/CFT correspondence \cite{Raju:2012zs} and supersymmetry in general \cite{Bastianelli:1999ab}. 
In particular, using a holographic approach, these correlators allow to describe the curvature gravitational perturbations   
in a pre-inflationary phase of the early universe characterized by strong gravity \cite{McFadden:2011kk,Bzowski:2011ab, 
Maldacena:2011nz}. They play a crucial role in the study of the non-gaussian contributions to such perturbations at the level of the 
bispectrum (via the $TTT$ correlator) and of the trispectrum 
(via the $TTTT$), while the power spectrum is determined by the $TT$ \cite{McFadden:2009fg}. In all the cases the calculation of the scalar perturbations, which are the most significant phenomenologically, involves fully 
traced 3-point functions of EMT's,  of field theories including scalars $\phi^J$ (with $J=1,2,...n_\phi$), vectors 
$A_\mu^I$ (with $I=1,2,...n_A$) and fermions $\psi^L$ ( with $L=1,2,...n_\psi$) in the virtual 
corrections.  All the fields in such theories are in the adjoint of the gauge group $SU(N)$. 

The mapping is described, on the dual field theory side, in terms of an effective (t'Hooft) coupling constant $g_{eff}=g_{YM}^2 N/M$, with $g_{eff} <<1$, and requires a large$-N$ limit. $M$ is a typical momentum scale, related to the typical momenta of the correlators, and necessary in order to make $g_{eff}$ dimensionless. This implies that the gauge coupling ($g_{YM}/M$) has to be very weak, allowing an expansion of the dual theory in such a variable, which can be arrested to zeroth order, i.e. with free fields.

It has been pointed out in \cite{McFadden:2009fg, McFadden:2011kk} that the small amplitude ($O(10^{-9}$)) and the nearly scale invariant characters of the measured power spectrum, wich shows deviations from scale invariance which are of $O(10^{-3})$, are indeed predicted by such models. In particular, the large-$N$ limit, which is necessary in order to predict the small amplitude of the power spectrum, requires $N\sim 10^4$, given that its scales as $1/N^2$ in holographic models. At the same time, the small violation of scaling invariance in the same power spectrum is controlled by the t'Hooft coupling $g_{eff}$, therefore requiring that this coupling has to be small as well.

These considerations allow to simplify drastically the computation of these correlators on the dual side, as we have mentioned. In particular, we are allowed to deal with simple dual theories in order to identify the leading behaviour of the perturbative correlators which appear in the holographic formula for the perturbations. 

Henceforth, the non-abelian character of the dual gauge theory becomes unessential if we work at leading order in $g_{YM}$, and the vector contributions are proportional to those of a free abelian theory. 
This implies that all our computations can be and are performed 
in simple free field theories of scalars, abelian vectors and fermions, with the scalars and the fermions taken as gauge singlets. This choice simplifies the notations and allows us to obtain the correct result, to be used in the holographic mapping, just by introducing a correction factor, which will be inserted at the end.

The explicit form of the mapping has been given in \cite{McFadden:2011kk, Bzowski:2011ab}, for the 
$\langle \zeta \zeta\zeta\rangle$ (bispectrum) correlator, with $\zeta$ describing the gauge invariant curvature perturbation of the 
gravitational metric, which is mapped to the $TTT$. The same (uncontracted) 3-T correlator determines the bispectrum of more complex 
3-point functions, $\langle\zeta \zeta \gamma\rangle$, $\langle\zeta \gamma\gamma\rangle$ and $\langle\gamma\gamma\gamma\rangle$, 
involving tensor perturbations ($\gamma$) \cite{Bzowski:2011ab}. 

A similar, though more involved, mapping between the trispectrum correlator  $\langle
\zeta \zeta\zeta\zeta\rangle$ and the $TTTT$ 4-point function is expected to hold. The explicit form of this mapping is not yet available, since it involves a direct extension of the holographic approach developed in 
\cite{McFadden:2011kk, Bzowski:2011ab}. 

Even in the absence, at least at the moment, of a suitable generalization of the holographic expressions given in \cite{McFadden:2011kk, Bzowski:2011ab}, it is clear that a complete determination of the trispectrum in holographic models, given its complexity, is a two-stage process. This
requires 1) the explicit derivation of the holographic relation 
which maps the $\langle\zeta\zeta\zeta\zeta\rangle$ to the 4-T correlator, followed 2) by an explicit computation of these higher point functions via the dual mapping. 

For instance, in the case of the bispectrum ($\langle\zeta\zeta\zeta\rangle$), the holographic analysis has been put forward in \cite{McFadden:2011kk}, followed later on by an 
explicit computation of the relevant 3-D correlators $TTT$ given in \cite{Bzowski:2011ab}.
 
Our goal in this work is to make one step forward in this program and present the explicit form of the $TTTT$ (fully traced)  correlator. The explicit form of the complete - uncontracted - correlator (which is of rank-8) is computationally very involved due to the higher 
rank tensor reductions and it will not be discussed here. 
  
At the same time we will proceed with an independent recomputation of the $TTT$ correlator in $D=3$, which has been investigated in 
\cite{Bzowski:2011ab, Maldacena:2011nz}. We anticipate that our analysis is in complete agreement with the result given in 
\cite{Bzowski:2011ab} for this correlator, and we will discuss the mapping between our approach and the one of \cite{Bzowski:2011ab}. This 
agreement allows to test our methods before coming to their generalization in the 4-T case. 

As we have already mentioned, the study of the 4-T correlator in $D=3$ is free of the complications present in their $D=4$
counterparts, which are affected by the scale anomaly and require renormalization. Checks of our computations have been performed at 
various levels. We secure the consistency of the result in the 4-T case by verifying the Ward identities which are expected to hold.

Our work is organized as follows. After a summary section in which we outline our definitions and conventions, we move to a computation of the general form 
of the $TTT$ in our approach, followed by a brief section in which we provide a mapping between our result and those of \cite{Bzowski:2011ab}.  The Ward identities 
for the 3- and 4-point case, which are essential in order to test the consistency of our computations, are discussed together in a single section. We then move to the perturbative determination of the 4-T. We have collected in several 
appendices some of the technical aspects related to the diagrammatic expansion, specific to the $D=3$ case.  

\section{The search for non-gaussian fluctuations} 
Clues on the physics of the very early universe come from the analysis of the primordial gravitational fluctuations, which leave an
imprint on the cosmic microwave background (CMB) and on the evolution of large scale structures (see \cite{Mukhanov:1990me, Bartolo:2010qu}). So far, 
the cosmological data have shown to be compatible with the gaussian character of such fluctuations, which implies that they can be expressed in terms of 2-point functions. In Fourier space they define the so called 
"power spectrum" ($\Delta(k)$). In this case 3-point correlators associated to such fluctuations vanish, together with all the 
correlators containing an odd number of these fields.

Measurements of the power spectra of perturbations are not able to answer questions concerning the evolution and the interactions of 
quantum fields which generate such fluctuations in the very early universe. In fact, inflation models with different fields 
and interactions can lead to power spectra which are quite similar. For this reason, there is a justified hope that it will be 
possible to unveil, through the identification of a non-gaussian behaviour of such fluctuations, aspects of the physics of inflation 
which otherwise would remain obscure \cite{Komatsu:2009kd}. Tests of a possible non-gaussian behaviour of such perturbations can be performed using several 
observational probes, including analysis of the CMB, large scale structures and weak lensing, just to mention 
a few. 

One important result \cite{Maldacena:2002vr} in the study of the non-gaussian behaviour of single field inflation was the proof that 
in these models such fluctuations are small and, for this reason, the possible experimental detection of significant 
non-gaussianities would allow to rule them out.

\subsection{Domain-Wall/Cosmology correspondence and gauge/gravity duality}

An interesting approach \cite{McFadden:2010vh, Bzowski:2011ab, McFadden:2009fg} which allows to merge the analysis of fluctuations 
and of their quantization with ideas stemming from gauge/gravity duality, has been developed in the last few years. 
These formulations allow to define a
correspondence between two bulk theories, describing cosmological and domain wall gravitational backgrounds, and hence between their boundary duals, which are described by appropriate 3-D field theories.  
The two bulk metrics are related by an  
analytic continuation. Once that a cosmological model is mapped into a 4-D domain wall model, 
gauge/gravity duality can be used to infer the structure of the correlators in the bulk using a corresponding field theory on the boundary. Such a theory is not conformal and can 
be described by a combination of scalar, fermion and spin-1 sectors, formulated as simple field theories in flat 3-D backgrounds. 

Scalar and tensor fluctuations in domain wall backgrounds can then be described in terms of correlators involving multiple insertions 
of EMT's, computed in ordinary perturbation theory.  These result can be mapped back to describe the 
correspondence between bulk and boundary in the usual cosmological context, by an analytic continuation of the boundary correlators.  

In this framework, 
one can derive holographic formulas which allow to describe a primordial phase of strong gravity just by weakly coupled perturbations in the dual theory. We will present below, to make our discussion self-contained, the explicit expressions of one of these relations, which are of direct relevance for our analysis. 

We also mention that, in the conformal case, the 3-T and 4-T correlators in scalar and fermion free field theories provide a realization 
of the bispectrum and of the trispectrum of gravitational waves in De Sitter space \cite{Maldacena:2011nz}.
Discussions of the conformal properties of 3- and 4-point functions of primordial fluctuations can be found in \cite{ Kehagias:2012pd, Antoniadis:2011ib}. 

\section{Field theory realizations}

The correlation functions that we intend to study will be computed in four free field theories, namely
a minimally coupled and a conformally coupled scalar, a fermion and a spin 1 abelian gauge field.

If the classical theory is described by the action $\mathcal{S}$, the energy-momentum tensor (EMT) of the system is obtained by coupling it to a curved 3-D background metric $g_{\mu\nu}$ (with $\mathcal{S}\to \mathcal{S}[g]$) and functionally differentiating the action with respect to it. The formalism is similar to the ordinary one in the case of a 4-D gravitational spacetime  
\beq \label{EMT}
T^{\mu\nu}(z) = -\frac{2}{\sqrt{g_z}}\frac{\delta\,\mathcal{S}}{\delta g_{\mu\nu}(z)}\, ,
\eeq
and for this reason we will be using greek indices, with the understanding that they will run from 1 to 3. We will also set $\textrm{det}\, g_{\mu\nu}(z)\equiv g_z$ for the determinant of the 3-D metric.  

In the quantum theory, let $\mathcal W[g]$ be the euclidean generating functional depending on the classical background,
\beq\label{Generating}
\mathcal W[g] = \frac{1}{\mathcal{N}} \, \int \, \mathcal D\Phi \, e^{-\, \mathcal{S}} \, ,
\eeq
where $\mathcal{N}$ is a normalization factor, and $\Phi$ denotes all the quantum fields of the theory except the metric.
$\mathcal W[g]$ generates both connected and disconnected correlators of EMT's, which are 1-particle reducible. For notational simplicity we prefer to use this generating functional of the Green's function of the theory, rather than $\log \mathcal W$ and its Legendre transform. It is implicitly understood that, in the perturbative expansion of the corresponding correlators, we will consider only the connected components. In a 1-loop analysis the issue of 1-particle reducibility does not play any relevant role, and hence the use of $\mathcal W$ will make the manipulations more transparent.

Then it follows from (\ref{EMT}) that the quantum average of the EMT in the presence of the background source is given by
\beq 
< T^{\mu\nu}(z) >_g = 
\frac{2}{\sqrt{g_z}}\frac{\delta\, \mathcal W}{\delta\, g_{\mu\nu}(z)} \, .
\eeq
where the subscript $g$ indicates the presence of a generic metric background. 
Otherwise, all the correlators which do not carry a subscript $g$, are intended to be written in the flat limit. 
It is understood that the metric is generic while performing all the functional derivatives and ordinary differentiations of the correlators, and that the limit of flat space is taken only at the end. 

As we have mentioned above, we focus our analysis on the determination of the complete 3-T correlator of free (euclidean) field theories of scalar, vector and fermions in 3 space dimensions and on the 4-T fully traced correlator, which we are now going to introduce.


The actions for the scalar ($S$) and the chiral fermion field ($CF$), are respectively given by 
\beqa
\mathcal{S}_{S}
&=&
\frac{1}{2} \, \int d^3 x \, \sqrt{g}\,
\bigg[g^{\mu\nu}\,\nabla_\mu\phi\,\nabla_\nu\phi - \chi \, R \,\phi^2 \bigg]\, ,\label{scalarAction}\\
\mathcal{S}_{CF}
&=&
\frac{1}{2} \, \int d^3 x \, V \, {V_a}^\rho\,
\bigg[\bar{\psi}\,\gamma^a\,(\mathcal{D}_\rho\,\psi)
- (\mathcal{D}_\rho\,\bar{\psi})\,\gamma^a\,\psi \bigg] \, . \label{FermionAction}
\eeqa
Here $\chi$ is the parameter corresponding to the term of improvement obtained by coupling $\phi^2$ to the 3-D scalar curvature $R$.
We will be concerned with two cases for the scalar Lagrangian, the minimally and the conformally coupled ones.  $\chi = 0$ describes a minimally coupled scalar ($MS$).
In three dimensions, for $\chi = 1/8$ one has a classically conformal invariant theory (i.e. one with an EMT whose trace
vanishes upon use of the equations of motion), which is the second case that we will consider (denoted with the "conformal scalar" subscript, or $CS$). As we have mentioned, the absence of conformal anomalies guarantees that for any of these theories, those operators which are classically traceless, remain such also at quantum level.

The other conformal field theory which we will be concerned with, is the one describing the Lagrangian of a free chiral fermion $(CF)$  on a curved metric background \footnote{ Notice that even if our analysis is euclidean, the 3-D case that we discuss can be mapped straightforwardly to the analogous Minkowski one in $D=2+1$ by a simple analytic continuation.}. 
Here ${V_a}^\rho$ is the vielbein and $V (= \sqrt{g})$ its determinant, needed to embed
the fermion in such background, with its covariant derivative $\mathcal{D}_\mu$ defined as
\beq
\mathcal{D}_\mu = \pd_\mu + \Gamma_\mu =
\pd_\mu + \frac{1}{2} \, \Sigma^{bc} \, {V_b}^\sigma \, \nabla_\mu\,V_{c\sigma} \, .
\eeq
with $\Sigma^{ab} = \frac{1}{4}\, [\gamma^a,\gamma^b]$ in the fermion case.
Using the integrability condition of the vielbein
\beq
\nabla_{\mu}\, V_{b\sigma} = \pd_{\mu}\, V_{b\sigma} - \Gamma^\lambda_{\mu\sigma}\, V_{b\lambda} 
= \Omega_{ab,\mu}\, {V^a}_\sigma \, ,
\eeq
where $\Omega_{ab,\mu}$ is the spin connection, this can be expressed as
\beq \label{SpinCon}
\Omega_{a b ,\mu} = 
\frac{1}{2}\, V^{\lambda}_{a}\left( \pd_{\mu} V_{b \lambda} - \pd_{\lambda}V_{b \mu} \right)
- \frac{1}{2}\, V^{\lambda}_{b} \left( \pd_\mu V_{a \lambda} - \pd_\lambda V_{a \mu} \right) 
+ \frac{1}{2} \, V^{\rho}_{a}\, V^{\lambda}_{b}\, V^{h}_{\mu}\, \left(\pd_{\lambda}V_{h \rho}- \pd_{\rho}V_{h \lambda} \right). \, 
\eeq
After an expansion we can rewrite (\ref{FermionAction}) as
\beq
\mathcal S_{CF} 
= \frac{1}{2}\, \int d^3 x\, V\, {V_a}^\rho\, \bigg[\bar\psi\, \gamma^a\, (\pd_\rho\, \psi) - (\pd_\rho\,\bar{\psi})\,\gamma^a\,\psi
\bigg]
+\frac{1}{16}\, \int d^3 x\, V\, {V_a}^\rho\, \bar\psi\,\{\gamma^a,[\gamma^b,\gamma^c]\}\, \psi\, \Omega_{bc,\rho}\, .
\eeq
The functional derivative with respect to the metric appearing in (\ref{EMT}) is expressed in terms of the vielbein as
\beq\label{SymmVielbein}
\frac{\delta}{\delta g_{\mu\nu}(z)} = 
\frac{1}{4}\, \left( {V_a}^\nu(z)\, \frac{\delta}{\delta V_{a\mu}(z)} + {V_a}^\mu(z)\, \frac{\delta}{\delta V_{a\nu}(z)}\right)\, ,
\eeq
so that the EMT for a fermion field is defined by
\footnote{It is known that the result quoted below in
Eq. (\ref{FermionEMT}) is obtained in the classical theory using Eq. (\ref{EMT}) and functionally differentiating with just the first term of Eq. (\ref{SymmVielbein}),
with no need for symmetrization. Indeed this is true only if we use the equations of motion, which is not allowed 
in the computation of the quantum interaction vertices derived from the fermionic action.}
\beq\label{SymmFerTEI}
T^{\mu\nu}_{CF}(z)
\stackrel{def}{\equiv} 
- \frac{1}{2\,V(z)}\bigg(V^{a\mu}(z) \, \frac{\delta}{\delta {V^a}_\nu(z)} + V^{a\nu}(z)\, \frac{\delta}{\delta {V^a}_\mu(z)}\bigg)\,
\mathcal S_{CF} \, .
\eeq
Finally, the action for the gauge field ($GF$) is given by
\beq
\mathcal{S}_{GF} = \mathcal{S}_M + \mathcal{S}_{gf} + \mathcal{S}_{gh}\, ,
\eeq
where the three contributions are the Maxwell (M), the gauge-fixing and the ghost actions
\beqa
\mathcal{S}_M    &=&   \frac{1}{4} \, \int d^3 x \, \sqrt{g} \, F^{\a\b} F_{\a\b}\, ,\\
\mathcal{S}_{gf} &=&   \frac{1}{2 \xi} \, \int d^3 x \, \sqrt{g} \, \left( \nabla_{\alpha}A^\alpha \right)^2\, ,\\
\mathcal{S}_{gh} &=& - \int d^3 x \, \sqrt{g}\, \partial^\a \bar{c} \, \partial_\a c\, .
\eeqa
The EMT's for the scalar and the fermion are
\beqa
T^{\mu\nu}_{S}
&=&
\nabla^\mu \phi \, \nabla^\nu\phi - \frac{1}{2} \, g^{\mu\nu}\,g^{\alpha\beta}\,\nabla_\alpha \phi \, \nabla_\beta \phi
+ \chi \bigg[g^{\mu\nu} \Box - \nabla^\mu\,\nabla^\nu + \frac{1}{2}\,g^{\mu\nu}\,R - R^{\mu\nu} \bigg]\, \phi^2 \,,
\label{ScalarEMT}\\
T^{\mu\nu}_{CF}
&=&
\frac{1}{4} \,
\bigg[ g^{\mu\rho}\,{V_a}^\nu + g^{\nu\rho}\,{V_a}^\mu - 2\,g^{\mu\nu}\,{V_a}^\rho \bigg]
\bigg[\bar{\psi} \, \gamma^{a} \, \left(\mathcal{D}_\rho \,\psi\right) -
\left(\mathcal{D}_\rho \, \bar{\psi}\right) \, \gamma^{a} \, \psi \bigg] \, ,
\label{FermionEMT}
\eeqa
while the energy-momentum tensor for the abelian gauge field is given by the sum
\beq
T^{\mu\nu}_{GF} = T^{\mu\nu}_M + T^{\mu\nu}_{gf} + T^{\mu\nu}_{gh}\, ,
\eeq
with
\beqa
T^{\mu\nu}_M
&=&
F^{\mu\a}{F^\nu}_{\a}  - \frac{1}{4}g^{\mu\nu}F^{\a\b}F_{\a\b} \, ,
\label{MaxwellEMT}
\\
T^{\mu\nu}_{gf}
&=&
\frac{1}{\xi}\{ A^\mu\nabla^\nu(\nabla_\r A^\r) + A^\nu\nabla^\mu(\nabla_\r A^\r ) -g^{\mu\nu}[ A^\r
\nabla_\r(\nabla_\s A^\s) + \frac{1}{2}(\nabla_\r A^\r)^2 ]\}\, ,
\label{GaugeFixEMT}
\\
T^{\mu\nu}_{gh}
&=&
g^{\mu\nu}\pd^{\r}\bar{c}\,\pd_{\r}c - \pd^\mu\bar{c}\, \pd^\nu c - \pd^\nu\bar{c}\, \pd^\mu c
\label{GhostEMT}\, .
\eeqa
The explicit expressions for the vertices involving one or more EMT's, which can be computed by functional differentiating the actions, 
have been collected in Appendix \ref{Vertices}. \\
We point out that in our computation of the contributions related to the gauge fields (GF), only the Maxwell
action $\mathcal{S}_M$ and the corresponding EMT, $T^{\mu\nu}_M$, are needed. One can check that there is a cancellation between the gauge-fixing and ghost contributions from $T_{gf}$ and $T_{gh}$. For those interested in a 
direct check of these results, we remark that the contributions generated from (\ref{ScalarEMT}) and those generated from (\ref{GhostEMT}) differ only by an overall sign factor, while the trilinear and quadrilinear vertices for the gauge-fixing part can be found in \cite{Coriano:2012wp}.

\section{Conventions and the structure of the correlators}
\begin{figure}[t]
\centering
\includegraphics[scale=0.8]{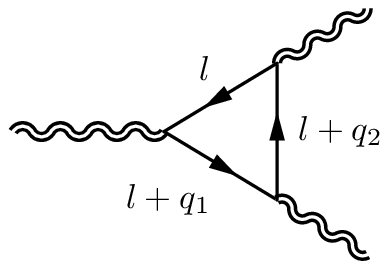}
\hspace{1cm}
\includegraphics[scale=0.8]{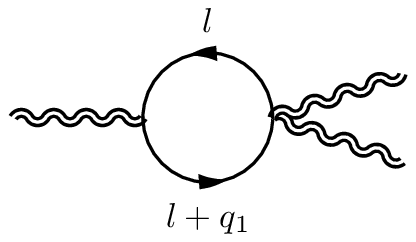}
\hspace{1cm}
\includegraphics[scale=0.8]{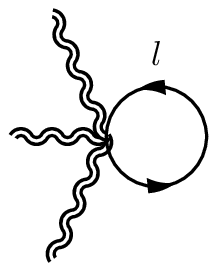}
\caption{Topologies appearing in the expansion of the $TTT$ correlator. Contributions involving coincident gravitons correspond to 
contact terms. \label{TTTtop} }
\end{figure}
We will be introducing two different notations for correlators involving the EMT's. The first is defined in terms of the symmetric 
$n-th$ order functional derivative of $\mathcal{W}$
\beqa \label{NPF}
< T^{\mu_1\nu_1}(x_1)...T^{\mu_n\nu_n}(x_n)> 
&=&
\bigg[\frac{2}{\sqrt{g_{x_1}}}...\frac{2}{\sqrt{g_{x_n}}} \,
\frac{\delta^n \mathcal{W}}{\delta g_{\mu_n\nu_n}(x_n) \ldots \delta g_{\mu_1\nu_1}(x_1)}\bigg]
\bigg|_{g_{\mu\nu} = \delta_{\mu\nu}} \nn \\
&=&
2^n\, \frac{\delta^n \mathcal{W}}{\delta g_{\mu_n\nu_n}(x_n) \ldots \delta g_{\mu_1\nu_1}(x_1)}\bigg|_{g_{\mu\nu} = 
\delta_{\mu\nu}} \, .
\eeqa
We will refer to this correlator as to a "symmetric" one. Notice that given the existence of an analytic continuation between the 3-D euclidean theory and the one in 2+1 dimensional spacetime, we will be referring to this vertex, for simplicity, as to the "n-graviton " vertex. 
 
This definition allows to leave the factor ${2}/{\sqrt{g}}$ outside of the actual differentiation in order to obtain symmetric expressions. We 
have denoted this correlator with a small angular brackets ($< \,\, >$) since these correlators include also 
contact terms.  Contact 
terms are easily identified in perturbation theory for bringing together at least two gravitons on the same spacetime point. Such terms are instead absent in the expression of correlation functions given by the (ordinary) expectation value of the product of $n$ EMT's, and which are denoted, in our case, with large angular brackets ($\langle\,\, \rangle$)  as in 
\beq
\langle T^{\mu_1\nu_1}(x_1) \, \ldots \, T^{\mu_n\nu_n}(x_n) \rangle =  \frac{1}{\mathcal N} \int \mathcal D \Phi \, 
T^{\mu_1\nu_1}(x_1) \, \ldots \, T^{\mu_n\nu_n}(x_n)  \, e^{-S} \bigg|_{g_{\mu\nu}=\delta_{\mu\nu}} \,.
\eeq
This second form of the correlator of EMT's will be referred to as "ordinary" or "genuine" $n$-point functions. 
It will also be useful to introduce the following notation to represent the functional derivative with respect to the background metric
\beqa
\label{funcder}
\left[f(x)\right]^{\muu\nuu\mud\nud\dots\mu_{n}\nu_{n}}(\xu,\xd,\dots,x_n) 
\equiv
\frac{\delta^n\, f(x)}{\delta g_{\mu_n\nu_n}(x_{n}) \, \ldots \, \delta g_{\mud\nud}(\xd) \, \delta g_{\muu\nuu}(\xu)}
\bigg|_{g_{\mu\nu}=\delta_{\mu\nu}} 
\eeqa
and take the flat spacetime limit at the end.
For later use we also define the notation with lower indices as
\beqa
\left[f(x)\right]_{\muu\nuu\ldots\mu_{n}\nu_{n}}(\xu,\xd,\dots,x_n) 
\equiv
\delta_{\mu_1 \alpha_1} \delta_{\nu_1 \beta_1}\, \ldots\, \delta_{\mu_n \alpha_n}\delta_{\nu_n \beta_n}\,
\left[f(x)\right]^{\alpha_1 \beta_1 \alpha_2 \beta_2 \dots \alpha_{n}\beta_{n}}(\xu,\xd,\dots,x_n)  \,.
\eeqa
With this definition a single functional derivative of the action in a correlation function is always equivalent, modulo a factor, to an insertion of 
a $T^{\mu\nu}$ in the flat limit, since 
\beq
 \left[\mathcal S \right]^{\mu_1 \nu_1}(x_1) \equiv \frac{\delta \mathcal S}{\delta g_{\mu_1 \nu_1}(x_1)} \bigg|_{g_{\mu\nu}=\delta_{\mu\nu}} =  -\frac{1}{2}T^{\mu_1 \nu_1}(x_1) \,.
 \eeq
We can convert the two graviton (greek) indices $(\mu_1,\nu_1)$ into a latin index $(s_1)$ by contracting with a polarization vector 
of generic polarization $s_1$ as in 
\beq
\left[\mathcal S\right]^{(s_1)} = -\frac{1}{2}T^{(s_1)}\equiv\left[\mathcal S \right]^{\mu_1\nu_1}\epsilon_{\mu_1\nu_1}^{(s_1) \, *},
\eeq
with $s_1\equiv \pm$, being the two helicities, as explained below. Moreover the symbol $T$ will be used, in a correlator, to denote 
the trace of the EMT,  $T \equiv {T^\mu}_\mu$.
We will also stack the two $(\mu_1\nu_1)$ indices, one on top of the other, for simplicity, after a contraction, as in 
\beq 
\left[\mathcal S \right]^{\mu_1}_{\mu_1}\equiv  \left[\mathcal S \right]^{\mu_1\nu_1}\, 
\delta_{\mu_1 \nu_1}\qquad \textrm{or}
\qquad  
\left[\mathcal S \right]^{\mu_1\mu_2}_{\mu_1\mu_2} \equiv  \left[\mathcal S \right]^{\mu_1\nu_1\mu_2\nu_2}
\delta_{\mu_1 \nu_1} \delta_{\mu_2 \nu_2} 
\eeq
in order to make the tensorial expressions more compact. 

With the definition of Eq. (\ref{NPF}) the expansion of the $TT$ correlator becomes
\bea
< T^{\muu\nuu}(x_1)T^{\mud\nud}(x_2) >
= 4\, \bigg[ \langle \left[\mathcal S \right]^{\muu\nuu}(\xu)\left[\mathcal S \right]^{\mud\nud}(\xd) \rangle - \langle 
[S]^{\muu\nuu\mud\nud}(\xu,\xd) \rangle \bigg] \,.
\eea
The last term on the right hand side of the equation above, which is a massless tadpole, is scheme dependent and can be easily 
removed by a local finite counterterm. For this reason it can be set to zero and then the $TT$ correlation function, obtained by 
differentiation of the generating functional, coincides with the quantum average of two energy momentum tensors
\bea
< T^{\muu\nuu}(x_1)T^{\mud\nud}(x_2) > &=& 4  \langle \left[\mathcal S \right]^{\muu\nuu}(\xu)\left[\mathcal S 
\right]^{\mud\nud}(\xd) \rangle  
= \langle T^{\muu\nuu}(x_1)T^{\mud\nud}(x_2) \rangle \,.
\eea
This is not true for higher order correlation functions of $n$-gravitons, as we are going to show in a while, where contact terms also appear. \\
For the $TTT$ correlator the functional expansion is given by
\bea
\label{3Tall}
< T^{\muu\nuu}(x_1)T^{\mud\nud}(x_2)T^{\mut\nut}(x_3) >
&=&
\langle T^{\muu\nuu}(\xu) T^{\mud\nud}(\xd)T^{\mut\nut}(\xt)\rangle 
\nn \\
&& \hspace{-25mm} 
-\, 4\, \bigg(\langle \left[\mathcal S\right]^{\muu\nuu\mud\nud}(\xu,\xd)\,T^{\mut\nut}(\xt)\rangle + 2\, \text{perm.} \bigg)
-\, 8\, \langle  \left[\mathcal S\right]^{\muu\nuu\mud\nud\mut\nut}(\xu,\xd,\xt)\rangle \bigg]   
\eea
whose right hand side is expressed in terms of ordinary 3-point correlators plus extra contact terms. The additional 
terms obtained by permutation are such to render symmetric the right hand side of the previous equation. 

We will present the expression of these contributions in the helicity basis for each sector (scalar, fermion and gauge field) in the 
sections below. Notice that the first term on the right hand side of Eq. (\ref{3Tall}) corresponds to an ordinary (genuine) 3-point function, whose connected component is given, at 1 loop, by the triangle diagram of Fig. \ref{TTTtop},
while the last term is a massless tadpole (see the third diagram in Fig. 
\ref{TTTtop}), which can be set to zero
\beq
\langle \left[\mathcal S\right]^{\muu\nuu\mud\nud\mut\nut}(\xu,\xd,\xt)\rangle=0. 
\eeq
In the 3-T case, contact terms have the topology of a bubble, and are generated by correlators containing insertions of the second functional derivatives of the action respect to the metric 
(such as in $\left[S\right]^{\mut\nut\muq\nuq}$). One of them is shown in the second figure of Fig. \ref{TTTtop}, the others can be 
obtained by a reparameterization of the external momenta. These bubble terms are characterized by the insertion of two graviton lines 
on the same vertex. \\
Moving to the 4-T case, a similar expansion holds and is given by
\beqa
< T^{\muu\nuu}(x_1)T^{\mud\nud}(x_2)T^{\mut\nut}(x_3)T^{\muq\nuq}(x_4)>
&=&
\langle T^{\muu\nuu}(x_1)T^{\mud\nud}(x_2)T^{\mut\nut}(x_3)T^{\muq\nuq}(x_4)\rangle
\nn \\
&& \hspace{-75 mm}
-\, 4\, \bigg[\langle \left[\mathcal S\right]^{\muu\nuu\mud\nud}(\xu,\xd)T^{\mut\nut}(\xt) T^{\muq\nuq}(\xq)\rangle 
+ 5 \, \text{perm.} \bigg] + 16\, \bigg[\langle \left[\mathcal S\right]^{\muu\nuu\mud\nud}(\xu,\xd)\left[\mathcal 
S\right]^{\mut\nut\muq\nuq}(\xt,\xq)\rangle + 2 \, \text{perm.} \bigg]
\nn \\
&& \hspace{-75mm}
- \, 8\, \bigg[\langle \left[\mathcal S\right]^{\muu\nuu\mud\nud\mut\nut}(\xu,\xd,\xt)T^{\muq\nuq}(\xq)\rangle + 3 \, \text{perm.} \bigg]
- \, 16\, \langle \left[\mathcal S\right]^{\muu\nuu\mud\nud\mut\nut\muq\nuq}(\xu,\xd,\xt,\xq)\rangle \, 
\label{4PF}
\eeqa
with 
\beq
\langle \left[\mathcal S\right]^{\muu\nuu\mud\nud\mut\nut\muq\nuq}(\xu,\xd,\xt,\xq)\rangle=0, 
\eeq
being a massless tadpole contribution. Notice that the left hand side and the right hand side are both symmetric amplitudes, as they 
should. In this case the perturbative expansion in the three fundamental sectors - scalars, vector and fermion - generates diagrams of box type 
for the first 4-T correlator on the right hand side of (\ref{4PF}), plus triangle, bubble and tadpole diagrams generated by the 
contact terms. The analysis of these contributions is more involved compared to the $TTT$ case, and will be performed in detail in 
the following sections. 
%

\section{Ward identities for the Green functions}

We proceed with a derivation of the relevant Ward identities satisfied by the 3- and 4-point functions of EMT's.

The diffeomorphism Ward identities are defined from the condition of general covariance on the generating functional $\mathcal W[g]$ 
\beq\label{masterWI0}
\nabla_{\nuu} < T^{\muu\nuu}(x_1) >_g
= \nabla_{\nuu} \bigg(\frac{2}{\sqrt{g_{x_1}}}\frac{\delta\mathcal W[g]}{\delta g_{\muu\nuu}(x_1)}\bigg)
= 0 \, .
\eeq
The Ward identities we are interested in are obtained by functional differentiation of Eq. (\ref{masterWI0}) and are given by
\bea
\pd_{\nuu} < T^{\muu\nuu}(x_1) T^{\mud\nud}(x_2) > &=& 0 \, , \label{WI2PFCoordinateFlat}
\eea
which shows that the two-point function is transverse, and by
\bea
\pd_{\nuu} < T^{\muu\nuu}(x_1)T^{\mud\nud}(x_2)T^{\mut\nut}(x_3) > 
&=& 
- 2\, \left[\Gamma^{\muu}_{\kappa\nuu}(x_1)\right]^{\mud\nud}(x_2) \langle T^{\kappa\nuu}(x_1)T^{\mut\nut}(x_3) \rangle
\nn \\
&&
- 2\, \left[\Gamma^{\muu}_{\kappa\nuu}(x_1)\right]^{\mut\nut}(x_3) \langle T^{\kappa\nuu}(x_1)T^{\mud\nud}(x_2) \rangle 
\, , \label{WI3PFCoordinateFlat} \\
\pd_{\nuu} < T^{\muu\nuu}(x_1)T^{\mud\nud}(x_2)T^{\mut\nut}(x_3)T^{\muq\nuq}(x_4) > 
&=&
- 2\, \bigg[\left[\Gamma^{\muu}_{\kappa\nuu}(x_1)\right]^{\mud\nud}(x_2)
 < T^{\kappa\nuu}(x_1) T^{\mut\nut}(x_3)T^{\muq\nuq}(x_4) >
\nn \\
&& \hspace{-55mm}
+ \big( 2 \leftrightarrow 3, 2 \leftrightarrow 4 \big) \bigg]
- 4\, \bigg[ \left[\Gamma^{\muu}_{\kappa\nuu}(x_1)\right]^{\mud\nud\mut\nut}(x_2,x_3)
 \langle T^{\kappa\nuu}(x_1)T^{\muq\nuq}(x_4) \rangle 
+ \big( 2 \leftrightarrow 4, 3 \leftrightarrow 4 \big) \bigg] \,  \label{WI4PFCoordinateFlat} 
\eea
for the 3- and 4-point functions.
The functional derivatives of the Christoffel symbol, obtained from the expansion of the covariant derivative which appear in previous equations, are explicitly given in Appendix \ref{Vertices}. 

Before moving to momentum space, we define the Fourier transform using the following conventions
\beqa
&&
\int \, d^3 x_1 \,d^3 x_2 \, \ldots \,d^3 x_n \, 
\langle T^{\mu_1\nu_1}(x_1) \, T^{\mu_2\nu_2}(x_2) \, \ldots \, T^{\mu_n\nu_n}(x_n)\rangle \,
e^{-i(k_1\cdot x_1 + k_2 \cdot x_2 + \ldots + k_n \cdot x_n)} = 
\nn \\
&& \hspace{40mm}
(2\pi)^3\,\delta^{(3)}(\vec{k_1}+ \vec{k_2} + \ldots + \vec{k_n})\,\langle T^{\mu_1\nu_1}(\vec{k_1}) \, T^{\mu_2\nu_2}(\vec{k_2}) \, 
\ldots \, T^{\mu_n\nu_n}(\vec{k_n})\rangle \, , 
\label{4PFMom}
\eeqa
where all the momenta are incoming. Similar conventions are chosen for the 2-, 3- and 4-point functions in all the expressions that 
follow. We will consider Fourier-transformed correlation functions with an implicit momentum conservation, i.e. we will omit the 
overall delta function. Tridimensional momenta in the perturbative expansions will be denoted as $\vec{k}$ with components $k_\mu$. The modulus of $\vec{k}$ will be simply denoted as $k$.

Going to momentum space the $TT$ correlator satisfies the simple Ward identity
\bea
k_{1\,\nuu} \langle T^{\muu\nuu}(\ku)T^{\mud\nud}(-\ku) \rangle = 0 \,,
\eea
while for the $TTT$ three-point function we obtain
\beqa
k_{1\,\nuu} < T^{\muu\nuu}(\ku) T^{\mud\nud}(\kd) T^{\mut\nut}(\kt) >
&=&
- k_3^{\muu} \langle T^{\mut\nut}(\kd)  T^{\mud\nud}(- \kd)  \rangle 
- k_2^{\muu} \langle T^{\mud\nud}(\kt) T^{\mut\nut}(- \kt) \rangle \nn \\
&&  \hspace{-1cm}
+ k_{3\, \nuu} \bigg[\delta^{\muu\nut} \langle T^{\nuu\mut}(\kd) T^{\mud\nud}(-\kd) \rangle
+ \delta^{\muu\mut} \langle T^{\nuu\nut}(\kd) T^{\mud\nud}(-\kd)  \rangle \bigg] 
\nn \\
&&  \hspace{-1cm}
+ k_{2\, \nuu} \bigg[\delta^{\muu\nud} \langle T^{\nuu\mud}(\kt) T^{\mut\nut}(-\kt)  \rangle
+ \delta^{\muu\mud}  \langle T^{\nuu\nud}(\kt) T^{\mut\nut}(-\kt) \rangle\bigg],
\eeqa
and finally, for the $TTTT$
\beqa
&&
k_{1\,\nuu} \,< T^{\muu\nuu}(\ku)T^{\mud\nud}(\kd)T^{\mut\nut}(\kt)T^{\muq\nuq}(\kq) > =
\bigg[- k_2^{\muu} \, < T^{\mud\nud}(\ku+\kd)T^{\mut\nut}(\kt)T^{\muq\nuq}(\kq) >
\nn \\
&+&
 k_{2\,\nuu}\, \bigg( \delta^{\muu\nud}\, < T^{\nuu\mud}(\ku+\kd)T^{\mut\nut}(\kt)T^{\muq\nuq}(\kq) >
+ \delta^{\muu\mud}\, < T^{\nuu\nud}(\ku+\kd)T^{\mut\nut}(\kt)T^{\muq\nuq}(\kq) >  \bigg)
\nn \\
&+&  \big( 2 \leftrightarrow 3, 2 \leftrightarrow 4\big) \bigg] 
+ \bigg[ 2\, k_{2\,\nuu}\, \bigg( 
\left[g^{\muu\mud}\right]^{\mut\nut}\, \langle T^{\nuu\nud}(\kq) T^{\muq\nuq}(-\kq) \rangle + 
\left[g^{\muu\nud}\right]^{\mut\nut}\, \langle T^{\nuu\mud}(\kq)T^{\muq\nuq}(-\kq) \rangle \bigg)
\nn \\
&+& \hspace{2mm}
 2\, k_{3\,\nuu}\, \bigg(
\left[g^{\muu\mut}\right]^{\mud\nud}\, \langle T^{\nuu\nut}(\kq) T^{\muq\nuq}(-\kq)  \rangle + 
\left[g^{\muu\nut}\right]^{\mud\nud}\, \langle T^{\nuu\mut}(\kq) T^{\muq\nuq}(-\kq)  \rangle \bigg)
\nn \\
&+&
 \bigg( k_2^{\nut} \delta^{\muu\mut} + k_2^{\mut} \delta^{\muu\nut}\bigg)\, \langle T^{\mud\nud}(\kq) T^{\mut\nut}(-\kq) \rangle
+ \bigg( k_3^{\nud} \delta^{\muu\mud} + k_3^{\mud} \delta^{\muu\nud}\bigg)\, \langle T^{\mut\nut}(\kq) T^{\mud\nud}(-\kq)  \rangle 
\nn \\
&+&
 \big( 2 \leftrightarrow 4, 3 \leftrightarrow 4 \big) \bigg] \,. \label{WI4PFMomentumFlat}
\eeqa
The functional derivatives of the metric tensor are computed using Eq. (\ref{funcder}) and given explicitly in Appendix
\ref{Vertices}.

For any conformal field theory in an odd dimensional spacetime the relation
\beq \label{TRACE}
g_{\mu\nu}\, < T^{\mu\nu} >_g = < {T^\mu}_\mu >_g = 0 \, 
\eeq
describes the invariance under scale transformations. This allows us to derive additional constraints on the fermion and on the conformally coupled scalar
correlators. Differentiating (\ref{TRACE}) with respect to the metric up to three times
and then performing the flat limit we obtain the three Ward identities
\beqa
< T(\vec{k_1})T^{\mud\nud}(\vec{-k_1})> &=& 0\, , \label{AWard2PF} \nn \\
< T(\vec{k_1})T^{\mud\nud}(\vec{k_2})T^{\mut\nut}(\vec{k_3}) > 
&=&  - 2 < T^{\mud\nud}(\vec{k_2})T^{\mut\nut}(\vec{-k_2})> - 2 < T^{\mud\nud}(\vec{k_3})T^{\mut\nut}(\vec{-k_3})>  \nn\\
< T(\vec{k_1})T^{\mud\nud}(\vec{k_2})T^{\mut\nut}(\vec{k_3})T^{\muq\nuq}(\vec{k_4}) > 
&=& 
-2\, < T^{\mud\nud}(\vec{k_2})T^{\mut\nut}(\vec{k_3})T^{\muq\nuq}(\vec{k_2}+\vec{k_3}) > \label{AWard3PF}
\nn \\
&& \hspace{-50mm}
-\,  2 \, < T^{\mud\nud}(\vec{k_3}+\vec{k_4})T^{\mut\nut}(\vec{k_3})T^{\muq\nuq}(\vec{k_4}) >
-2\, < T^{\mud\nud}(\vec{k_2})T^{\mut\nut}(\vec{k_2}+\vec{k_4})T^{\muq\nuq}(\vec{k_4}) >
\, . \label{AWard4PF}
\eeqa
Tracing the last two equations with respect to the remaining two and three couples of indices respectively we get the constraints
\beqa
< T(\vec{k_1})T(\vec{k_2})T(\vec{k_3})> &=& 0 \, ,
\nn \\
< T(\vec{k_1})T(\vec{k_2})T(\vec{k_3})T(\vec{k_4}) > &=& 0
\label{TraceConstraints}
\eeqa
valid in the conformal case.

\section{Computation of $TTT$}

We begin this section recalling the results for the two-point function $TT$ in $D=3$, which takes the form 
\bea
\langle T^{\mu \nu}(\vec{k}) T^{\alpha \beta}(-\vec{k}) \rangle = 
A(k) \Pi^{\mu\nu\alpha\beta}(\vec{k}) + B(k) \pi^{\mu\nu}(\vec{k}) \pi^{\alpha\beta}(\vec{k})\, ,
\eea
where $\pi^{\mu\nu}(\vec{k})$ is a transverse projection tensor
\bea
\pi^{\mu\nu}(\vec{k}) = \delta^{\mu\nu} - \frac{k^{\mu} k^{\nu}}{k^2} \,,
\label{proj}
\eea
while $ \Pi^{\mu\nu\alpha\beta}(\vec{k})$ is transverse and traceless
\bea
\Pi^{\mu\nu\alpha\beta}(\vec{k}) = \frac{1}{2} \bigg[ \pi^{\mu\alpha}(\vec{k}) \pi^{\nu\beta}(\vec{k}) + 
\pi^{\mu\beta}(\vec{k}) \pi^{\nu\alpha}(\vec{k}) - \pi^{\mu\nu}(\vec{k}) \pi^{\alpha\beta}(\vec{k}) \bigg] \,.
\eea
The coefficients $A(k)$ and $B(k)$ for the minimal scalar case (MS), the conformally coupled scalar (CS), 
the gauge field (GF) and for the chiral fermion case (CF) are given by
\bea
&& 
A^{MS}(k) =  A^{CS}(k) =  A^{GF}(k) =  \frac{k^3}{256}\, , 
\qquad A^{CF}(k) =  \frac{k^3}{128} \,, \\
&& 
B^{MS}(k) =  B^{GF}(k) =  \frac{k^3}{256}\,, 
\qquad B^{CS}(k) = B^{CF}(k) = 0 \, .
\eea

At this point we proceed with the computation of the $TTT$ correlator. The contributions introduced in 
Eq. (\ref{3Tall}) are built out of the vertices listed in the Appendix \ref{Vertices} and computed in terms of the usual tensor-to-scalar reductions of the loop integrals in $D=3$. The computations are finite at any stage and require only a removal of the massless tadpoles. Given the complex structure of the tensor result for the 3-T case, here we prefer to give its helicity projections instead of 
presenting it in an expansion on a tensor basis. We follow the same approach presented in \cite{Bzowski:2011ab}. \\
We define the helicity tensors as usual as
\bea
\epsilon^{(s)}_{\mu \nu}(\vec{k}) = \epsilon^{(s)}_{\mu}(\vec{k}) \, \epsilon^{(s)}_{\nu}(\vec{k}) \qquad \mbox{with} \quad s = \pm 1
\eea
where $\epsilon^{(s)}_{\mu}$ is the polarization vector for a spin 1 in $D=3$. They satisfy the standard orthonormality, traceless and transverse conditions
\bea
\epsilon^{(s)}_{\mu \nu}(\vec{k}) \epsilon^{(s') \, \, \mu \nu \, \, *}(\vec{k}) = \delta^{s s'} \,, \qquad \delta^{\mu \nu} \epsilon^{(s)}_{\mu \nu}(\vec{k}) = 0 \,, \qquad k^{\mu}  \epsilon^{(s)}_{\mu \nu}(\vec{k}) = 0 \,.
\eea
We consider a particular realization of the helicity basis choosing, without loss of generality, the three incoming momenta $\ku, \kd$ and $\kt$ to lie in $(x-z)$ plane
\bea
k_i^{\mu} = k_i (\sin \theta_i , 0, \cos \theta_i)\, , 
\eea
with the angles completely determined from the kinematical invariants as
\bea
&& \cos \theta_1 = 1 \,, \qquad   \cos \theta_2 = \frac{1}{2 k_1 k_2} (k_3^2-k_1^2-k_2^2) \,, \qquad \cos \theta_3 = \frac{1}{2 k_1 k_3}(k_2^2 - k_1^2 - k_3^2) \,, \nn \\
&& \sin \theta_1 = 0  \,, \qquad   \sin \theta_2 = \frac{\lambda(k_1,k_2,k_3)}{2 k_1 k_2} \,,  \qquad \sin \theta_3 = - \frac{\lambda(k_1,k_2,k_3)}{2 k_1 k_3} \,
\eea
and where 
\bea
\lambda^2(q_1,q_2,q_3) = (- q_1 + q_2 + q_3)(q_1 - q_2 + q_3) ( q_1 + q_2 - q_3)(q_1 + q_2 + q_3) \,.
\eea
Then the helicity tensors are explicitly given by
\bea
\epsilon^{(s)}_{\mu \nu}(\vec{k}) = \frac{1}{2} \left(
\begin{array}{ccc}
\cos^2 \theta_i & i s \cos \theta_i & - \sin \theta_i \cos \theta_i \\
i s  \cos \theta_i & -1 & - i s \sin \theta_i \\
- \sin \theta_i \cos \theta_i & - i s \sin \theta_i & \sin^2 \theta_i
\end{array}
\right) \,.
\eea 
Notice that $\epsilon_{\mu\nu}^{(s) \, *}(\vec{k}) = \epsilon_{\mu\nu}^{(s)}(- \vec{k})$, which turns useful in comparing our results with those of \cite{Bzowski:2011ab}. Notice also the different normalization of the polarization tensor $\epsilon_{\mu\nu}^{(s)}(\vec{k})$ used by us respect to \cite{Bzowski:2011ab}, by a factor of $\frac{1}{\sqrt{2}}$, which should be kept into account when comparing the results of each $\pm$ helicity projection. 
We are now ready to present in the following sections the $TTT$ correlator for the minimal and the conformally coupled scalar, the 
gauge field and the chiral fermion. All the other missing helicity amplitudes can be obtained from those given here by parity 
transformations or momentum permutations.

\subsection{Minimally coupled scalar}

We list the results for the $TTT$ correlator with a minimal scalar running in the loop.
They correspond to two of the three fundamental topologies appearing in the Fig. \ref{TTTtop}.  The different contributions in Eq. (\ref{3Tall}) can be contracted with polarization 
tensors in order to extract the $\pm$ helicity amplitudes and the trace parts ($T\equiv T^\mu_\mu$). \\
The ordinary 3-point amplitudes are expressed only in terms of the Euclidean length of the external vectors and no relative angles. They are given by
\bea
\langle T(\ku) T(\kd)T(\kt) \rangle_{MS}
&=&  
-\frac{1}{128} \bigg\{ k_1^3+\left(k_2+k_3\right) k_1^2+\left(k_2-k_3\right)^2 k_1+\left(k_2+k_3\right) \left(k_2^2+k_3^2\right) 
\bigg\} \,, \nn\\
\langle T(\ku)T(\kd) T^{(+)}(\kt) \rangle_{MS} 
&=& 
 \frac{1}{1024 k_3^2 (k_1+k_2+k_3)} (k_1 - k_2 - k_3) (k_1 + k_2 - k_3) (k_1 - k_2 + k_3)
\bigg\{ 3 k_1^3 \nn \\
&+& 
\left(5 k_2+6 k_3\right) k_1^2+\left(5 k_2^2+4 k_3 k_2+3 k_3^2\right) k_1
+ 3 k_2 \left(k_2+k_3\right)^2 \bigg\} \,, \nn \\
\langle T(\ku) T^{(+)}(\kd) T^{(-)}(\kt) \rangle_{MS} 
&=& 
-\frac{\left(k_1+k_2-k_3\right)^2 \left(k_1-k_2+k_3\right)^2}{4096 k_2^2 k_3^2} 
\bigg\{ 5 k_1^3-\left(k_2^2+4 k_3 k_2+k_3^2\right) k_1+k_2^3+k_3^3\bigg\} \nn
\eea
\bea
\langle T(\ku) T^{(+)}(\kd) T^{(+)}(\kt) \rangle_{MS} 
&=& 
-\frac{\left(-k_1+k_2+k_3\right)^2}{4096 k_2^2 k_3^2 \left(k_1+k_2+k_3\right)^2} 
\bigg\{ k_3^7+\left(3 k_1+4
   k_2\right) k_3^6+2 \left(k_1^2+6 k_2 k_1+3 k_2^2\right)
   k_3^5 \nn \\
&+& \left(3 k_1^3+16 k_2 k_1^2+21 k_2^2 k_1+5 k_2^3\right)
   k_3^4+\left(17 k_1^4+36 k_2 k_1^3+38 k_2^2 k_1^2-8 k_2^3 k_1+5
   k_2^4\right) k_3^3 \nn \\
&+& \left(k_1+k_2\right)^2 \left(29 k_1^3+14
   k_2 k_1^2+9 k_2^2 k_1+6 k_2^3\right) k_3^2+4
   \left(k_1+k_2\right)^3 \left(5 k_1^3+k_2 k_1^2+k_2^3\right)
   k_3 \nn \\
&+& \left(k_1+k_2\right)^4 \left(5 k_1^3-k_2^2
   k_1+k_2^3\right) \bigg\} \nn\,,
\eea
\bea
\langle T^{(+)}(\ku) T^{(+)}(\kd) T^{(+)}(\kt) \rangle_{MS} 
&=&
-\frac{\left(k_1+k_2-k_3\right) \left(k_1-k_2+k_3\right)
   \left(-k_1+k_2+k_3\right)} {16384 k_1^2 k_2^2 k_3^2 \left(k_1+k_2+k_3\right)^3} 
 \bigg\{ 3 \left(k_1+k_2+k_3\right)^9 \nn \\
&-& 7
   \left(k_2 k_3+k_1 \left(k_2+k_3\right)\right)
   \left(k_1+k_2+k_3\right)^7+5 k_1 k_2 k_3
   \left(k_1+k_2+k_3\right)^6-64 k_1^3 k_2^3 k_3^3 \bigg\} \,, \nn 
\eea
\bea
\langle T^{(+)}(\ku) T^{(+)}(\kd) T^{(-)}(\kt) \rangle_{MS} 
&=& 
\frac{\left(k_1-k_2-k_3\right) \left(k_1+k_2-k_3\right)^3
   \left(k_1-k_2+k_3\right) } {16384 k_1^2 k_2^2 k_3^2 \left(k_1+k_2+k_3\right)}
\bigg\{ 3 k_3^5 + 4 \left(k_1+k_2\right) k_3^4 \nn \\
&+& \left(k_1^2+k_2^2\right) k_3^3 + \left(k_1^3+k_2^3\right) k_3^2 
+ 4 \left(k_1+k_2\right){}^2 \left(k_1^2 - k_2 k_1 +k_2^2\right) k_3 \nn \\
&+& 
\left(k_1+k_2\right)^3 \left(3 k_1^2 - k_2 k_1 + 3 k_2^2\right)\bigg\} \,. 
\eea
The helicity projections of the contact terms are
\bea
\langle T(\ku) [\mathcal S]^{\mu_2 \mu_3}_ {\mu_2 \mu_3}(\kd,\kt) \rangle_{MS} 
&=& 
\frac{k_1^3}{256} \, , \nn \\
\langle T^{(s_1)}(\ku) [\mathcal S]^{\mu_2 \mu_3}_ {\mu_2 \mu_3}(\kd,\kt) \rangle_{MS} &=& 0 \, , \nn \\
\langle T(\ku) [\mathcal S]^{\mu_2 \, (s_3)}_ {\mu_2}(\kd,\kt) \rangle_{MS} 
&=& 
-  \frac{k_1 \, \lambda^2(k_1,k_2,k_3)}{4096 \, k_3^2} \, ,\nn  \\
\langle T(\ku) [\mathcal S]^{(s_2) \, (s_3)}(\kd,\kt) \rangle_{MS}
&=& 
- \frac{k_1 \, \lambda^2(k_1,k_2,k_3)}{4096 \, k_2^2 \, k_3^2}\bigg\{ - k_1^2 + k_2^2 + k_3^2 + 2 s_2 s_3 \, k_2 k_3\bigg\}\, ,
\nn
\eeqa
\beqa
\langle T^{(s_1)}(\ku) [\mathcal S]^{\mu_2 \, (s_3)}_ {\mu_2}(\kd,\kt) \rangle_{MS}
&=& 
\frac{k_1}{16384 \, k_3^2} \bigg\{ 
k_1^4 + k_2^4 + k_3^4 - 2 k_1^2 k_2^2 - 2 k_2^2 k_3^2 + 6 k_1^2 k_3^2 \nn \\
&&+ 
4 s_1 s_3 \, k_1 k_3  (k_1^2 -k_2^2 + k_3^2) \bigg\} \nn
\eeqa
\beqa
\langle T^{(s_1)}(\ku) [\mathcal S]^{(s_2) \, (s_3)}(\kd,\kt) \rangle_{MS}
&=& 
-\frac{k_1 \, \lambda^2(k_1,k_2,k_3)}{16384 \, k_2^2 k_3^2} \bigg\{ k_1^2 + k_2^2 + k_3^2  
+ 2 s_1 s_2 \, k_1 k_2 + 2 s_1 s_3 \, k_1 k_3 \nn \\
&&+  
2 s_2 s_3 \, k_2 k_3 \bigg\}\, . 
\eea
\subsection{Conformally coupled scalar}
In the case of a conformally coupled scalar the helicity amplitudes of the 3-point correlators are 
\bea
\langle  T(\ku) T(\kd) T(\kt) \rangle_{CS} &=& 0 \,, \nn \\
\langle T(\ku) T(\kd) T^{(+)}(\kt) \rangle_{CS} &=& 0 \,, \nn \\
\langle T(\ku) T^{(s_2)}(\kd) T^{(s_3)}(\kt) \rangle_{CS} 
&=& 
\frac{1}{4} \langle T(\ku) T^{(s_2)}(\kd) T^{(s_3)}(\kt) \rangle_{CF} \,, \nn \\
\langle T^{(s_1)}(\ku) T^{(s_2)}(\kd) T^{(s_3)}(\kt) \rangle_{CS}
&=&  
\langle T^{(s_1)}(\ku) T^{(s_2)}(\kd) T^{(s_3)}(\kt) \rangle_{MS} \,, 
\eea
while the expressions of the contact terms take the form 
\bea
\langle T(\ku) [\mathcal S]^{\mu_2 \mu_3}_ {\mu_2 \mu_3}(\kd,\kt) \rangle_{CS} &=& 0 \,, \nn \\
\langle T(\ku) [\mathcal S]^{\mu_2 \mu_3}_ {\mu_2 \mu_3}(\kd,\kt) \rangle_{CS} &=& 0 \,, \nn \\
\langle T(\ku) [\mathcal S]^{\mu_2 \, (s_3)}_ {\mu_2}(\kd,\kt) \rangle_{CS} &=& 0 \,, \nn \\
\langle T(\ku) [\mathcal S]^{(s_2) \, (s_3)}(\kd,\kt) \rangle_{CS} &=& 0 \,, \nn \\
\langle T^{(s_1)}(\ku) [\mathcal S]^{\mu_2 \, (s_3)}_ {\mu_2}(\kd,\kt) \rangle_{CS} 
&=&  
 \langle T^{(s_1)}(\ku) [\mathcal S]^{\mu_2 \, (s_3)}_ {\mu_2}(\kd,\kt) \rangle_{MS} \,, \nn   \\
\langle T^{(s_1)}(\ku) [\mathcal S]^{(s_2) \, (s_3)}(\kd,\kt) \rangle_{CS}
&=& 
\langle T^{(s_1)}(\ku) [\mathcal S]^{(s_2) \, (s_3)}(\kd,\kt) \rangle_{MS} \,.
\eea
\subsection{Gauge field}

Moving to the gauge contributions, a computation of the ordinary 3-point functions gives 
\bea
\langle T(\ku) T(\kd) T(\kt) \rangle_{GF}
&=& 
-\frac{1}{128} \bigg\{ -3 k_1^3+\left(k_2+k_3\right)
   k_1^2+\left(k_2-k_3\right)^2
   k_1-\left(k_2+k_3\right) \left(3 k_2^2-4 k_3 k_2+3 k_3^2\right) \bigg\}
\, , \nn
\eea
\bea
\langle T(\ku) T(\kd)T^{(+)}(\kt) \rangle_{GF} 
&=&
\frac{1}{1024 \,
  k_3^2 \left(k_1+k_2+k_3\right)} \left(k_1-k_2-k_3\right)
   \left(k_1+k_2-k_3\right) \left(k_1-k_2+k_3\right)
   \bigg\{5 k_1^3 \nn \\
&&+ \left(11 k_2+10 k_3\right)
   k_1^2+\left(11 k_2^2+12 k_3 k_2+5 k_3^2\right)
   k_1+5 k_2 \left(k_2+k_3\right)^2\bigg\} 
 \,, \nn 
\eea
\bea
\langle T(\ku)T^{(+)}(\kd) T^{(-)}(\kt) \rangle_{GF} 
&=&
\frac{\left(k_1^2- (k_2-k_3 )^2\right)^2 }{4096 \, k_2^2
   k_3^2} 
   \bigg\{7 k_1^3-\left(3 k_2^2+4 k_3 k_2+3
   k_3^2\right) k_1+k_2^3+k_3^3\bigg\} \,,  \nn \\
\langle T(\ku) T^{(+)}(\kd) T^{(+)}(\kt) \rangle_{GF} 
&=& 
\frac{\left(-k_1+k_2+k_3\right)^2}{4096 \, k_2^2 k_3^2
   \left(k_1+k_2+k_3\right)^2} 
   \bigg\{7 k_1^7 + 28 \left(k_2+k_3\right) k_1^6  \nn \\
&&+ 
\left(39 k_2^2+88 k_3 k_2 + 39 k_3^2\right) k_1^5 + \left(k_2+k_3\right)
   \left(17 k_2^2+71 k_3 k_2+17 k_3^2\right)
   k_1^4 \nn \\
&&-
\left(k_2+k_3\right)^2 \left(7 k_2^2-34 k_3 k_2 + 7 k_3^2\right) k_1^3 
- 2 \left(k_2+k_3\right)^3 \left(3 k_2^2-5 k_3 k_2+3 k_3^2\right) k_1^2  \nn \\
&&+ 
\left(k_2^6+4 k_3 k_2^5 + 7 k_3^2 k_2^4 + 40 k_3^3 k_2^3 + 7 k_3^4 k_2^2+4 k_3^5 k_2+k_3^6\right) k_1  \nn \\
&&+
\left(k_2+k_3\right)^5 \left(k_2^2-k_3 k_2+k_3^2\right) \bigg\}\, , \nn 
\eea
\bea
\langle T^{(+)}(\ku) T^{(+)}(\kd) T^{(+)}(\kt) \rangle_{GF} 
&=&
-\frac{\left(k_1+k_2-k_3\right) \left(k_1-k_2+k_3\right)
   \left(-k_1+k_2+k_3\right)} {16384 \, k_1^2 k_2^2 k_3^2 \left(k_1+k_2+k_3\right)^3} 
 \bigg\{ 3 \left(k_1+k_2+k_3\right)^9 \nn \\
&&- 7
   \left(k_2 k_3+k_1 \left(k_2+k_3\right)\right)
   \left(k_1+k_2+k_3\right)^7+5 k_1 k_2 k_3
   \left(k_1+k_2+k_3\right)^6-64 k_1^3 k_2^3 k_3^3 \nn \\
&&-  4  \left(k_1+k_2+k_3\right)^6 \left(k_1^3+k_2^3+k_3^3\right)
\bigg\}\, , \nn 
\eea
\bea
\langle T^{(+)}(\ku) T^{(+)}(\kd) T^{(-)}(\kt) \rangle_{GF} 
&=&
\frac{\left(k_1-k_2-k_3\right) \left(k_1+k_2-k_3\right)^3
   \left(k_1-k_2+k_3\right) } {16384 \, k_1^2 k_2^2 k_3^2 \left(k_1 + k_2 + k_3\right)}
\bigg\{ 3 k_3^5+4 \left(k_1+k_2\right) k_3^4 \nn \\
&&+
\left(k_1^2+k_2^2\right) k_3^3 + \left( k_1^3 + k_2^3 \right) k_3^2 
+ 4 \left(k_1+k_2\right){}^2 \left(k_1^2-k_2 k_1+k_2^2\right) k_3 	\nn \\
&& +
\left(k_1+k_2\right)^3 \left(3 k_1^2-k_2k_1+3 k_2^2\right) - 4 \left(k_1+k_2+k_3\right)^2 \left(k_1^3+k_2^3+k_3^3\right)
\bigg\}
 \,,
\eea
while the helicity projections of the contact terms are
\bea
\langle T(\ku) [\mathcal S]^{\mu_2 \mu_3}_ {\mu_2 \mu_3}(\kd,\kt) \rangle_{GF} &=& \frac{3 \, k_1^3}{256} \,, \nn \\
\langle T^{(s_1)}(\ku) [\mathcal S]^{\mu_2 \mu_3}_ {\mu_2 \mu_3}(\kd,\kt) \rangle_{GF} &=& 0 \,, \nn \\
\langle T(\ku) [\mathcal S]^{\mu_2 \, (s_3)}_ {\mu_2}(\kd,\kt)\rangle_{GF} 
&=& 
- \frac{3 \, k_1 \, \lambda^2(k_1,k_2,k_3)}{4096 \, k_3^2} \,, \nn \\
\langle T(\ku) [\mathcal S]^{(s_2) \, (s_3)}(\kd,\kt) \rangle_{GF} 
&=& 
 \frac{k_1^3}{2048 \, k_2^2 \,  k_3^2} \left\{ k_1^4+k_2^4+k_3^4  - 2 k_1^2 k_2^2  - 2 k_1^2 k_3^2  + 6 k_2^2 k_3^2 
\right. \nn \\
&&+
\left.
4\, k_2 k_3  s_2 s_3   \left( k_2^2 + k_3^2 -  k_1^2  \right) \right\}\, , \nn \\
\langle T^{(s_1)}(\ku) [\mathcal S]^{\mu_2 \, (s_3)}_ {\mu_2}(\kd,\kt) \rangle_{GF} 
&=&  
 \frac{3 \, k_1}{16384 \, k_3^2} \bigg\{ 
k_1^4 + k_2^4 + k_3^4 - 2 k_1^2 k_2^2 - 2 k_2^2 k_3^2 + 6 k_1^2 k_3^2 \nn \\
&&+ 
4 s_1 s_3 \, k_1 k_3  (k_1^2 -k_2^2 + k_3^2) \bigg\}\, , \nn \\
\langle T^{(s_1)}(\ku) [\mathcal S]^{(s_2) \, (s_3)}(\kd,\kt) \rangle_{GF} &=& 0 \,.
\eea
\subsection{Chiral fermion}
The analysis can be repeated in the fermion sector. In this case we obtain  
\bea
\langle T(\ku) T(\kd) T(\kt) \rangle_{CF} 
&=& 0 \,, \nn \\ 
\langle T(\ku) T(\kd) T^{(+)}(\kt) \rangle_{CF}
&=& 0 \,, \nn \\
\langle T(\ku) T^{(+)}(\kd) T^{(-)}(\kt) \rangle_{CF} 
&=& 
- \frac{k_2^3 + k_3^3}{1024 \, k_2^2 k_3^2} \bigg\{ k_1^2 - \left( k_2 - k_3 \right)^2 \bigg\}^2 \,, \nn \\
\langle T(\ku) T^{(+)}(\kd) T^{(+)}(\kt) \rangle_{CF} 
&=& 
- \frac{k_2^3 + k_3^3}{1024 \, k_2^2 k_3^2} \bigg\{ k_1^2 - \left( k_2 + k_3 \right)^2 \bigg\}^2  \,, \nn \\
\langle T^{(+)}(\ku) T^{(+)}(\kd) T^{(+)}(\kt) \rangle_{CF}
&=& 
-  \frac{\left(k_1+k_2-k_3\right) \left(k_1-k_2+k_3\right)
\left(-k_1+k_2+k_3\right)}{4096 \, k_1^2 k_2^2 k_3^2 \left(k_1+k_2+k_3\right)^3}
\bigg\{ 32 k_1^3 k_2^3 k_3^3 \nn \\
&& + \left(k_1+k_2+k_3\right)^9  -2 \left(k_2 k_3+k_1 \left(k_2+k_3\right)\right) \left(k_1+k_2+k_3\right)^7 \nn \\
&& + k_1 k_2 k_3 \left(k_1+k_2+k_3\right)^6 \bigg\}\, , \nn \\
\langle T^{(+)}(\ku) T^{(+)}(\kd) T^{(-)}(\kt) \rangle_{CF} 
&=& 
 \frac{\left(k_1-k_2-k_3\right) \left(k_1+k_2-k_3\right)^3 \left(k_1-k_2+k_3\right)}{4096\, k_1^2 k_2^2 k_3^2
\left(k_1+k_2+k_3\right)} \bigg\{ k_3^5+\left(k_1+k_2\right) k_3^4 \nn \\
&& - k_1 k_2 k_3^3 +\left(k_1+k_2\right)^2 \left(k_1^2-k_2 k_1+k_2^2\right) k_3+\left(k_1+k_2\right)^3 \left(k_1^2+k_2^2\right)\bigg\}\, ,
\eea
for the 3-point functions, while the helicity projections of the contact terms are
\bea
\langle T(\ku) [\mathcal S]^{\mu_2 \mu_3}_ {\mu_2 \mu_3}(\kd,\kt) \rangle_{CF} &=& 0 \,, \nn \\
\langle T^{(s_1)}(\ku) [\mathcal S]^{\mu_2 \mu_3}_ {\mu_2 \mu_3}(\kd,\kt) \rangle_{CF} &=& 0 \,, \nn \\
\langle T(\ku) [\mathcal S]^{\mu_2 \, (s_3)}_ {\mu_2}(\kd,\kt) \rangle_{CF} &=& 0 \,, \nn \\
\langle T(\ku) [\mathcal S]^{(s_2) \, (s_3)}(\kd,\kt) \rangle_{CF} &=&  0  \,, \nn \\
\langle T^{(s_1)}(\ku) [\mathcal S]^{\mu_2 \, (s_3)}_ {\mu_2}(\kd,\kt) \rangle_{CF} &=& 0 \,, \nn \\
\langle T^{(s_1)}(\ku) [\mathcal S]^{(s_2) \, (s_3)}(\kd,\kt) \rangle_{CF} 
&=& 
- \frac{3 \, k_1 \, \lambda^2(k_1,k_2,k_3)}{32768 \, k_2^2 k_3^2} \bigg\{ k_1^2 + k_2^2 + k_3^2
+ 2 s_1 s_2 \, k_1 k_2 + 2 s_1 s_3 \, k_1 k_3 \nn \\
&&+  
2 s_2 s_3 \, k_2 k_3
\bigg\} \,. 
\eea

\subsection{Multiplicites in the non-abelian case} 
As we have mentioned in the introduction, the expressions of the correlation functions in the small t'Hooft limit in the dual theory can be obtained from the results presented in the previous sections, which are computed for simple free field theories with gauge singlet fields, by introducting some appropriate overall factors. The prescription is to introduce a factor $(N^2-1)$ in front of each of our correlators (and contact terms), with the addition of multiplicity factors $n_\phi, n_\phi', n_A$ and $n_\psi$ in each sector. These corresponds to the multiplicites of the conformal scalars, minimally coupled scalars, gauge fields and fermions, respectively. For instance, the scalar $\langle TTT \rangle$ correlator is obtained with the replacements 
\beqa
\label{repper}
\langle TTT \rangle&\rightarrow& (N^2-1)\left( n_\psi \langle  TTT \rangle_{CF} +n_\phi \langle  TTT \rangle_{CS} +n_\phi' \langle TTT \rangle_{MS} + n_A \langle TTT \rangle_{GF}\right),
 \eeqa
 and similarly for all the others. 
 Notice that this results is an exact one. It reproduces, in leading order in the gauge coupling, what one expects for these correlators in a non-abelian gauge theory. The large-$N$ limit, in this case, is performed by replacing the color factor $N^2-1$ in front with $N^2$.

%
%
\section{Mapping of our result to the holographic one}
In holographic cosmology, as we have briefly pointed out in the previous sections, the gravitational cosmological perturbations are expressed in terms of correlators of field theories living on its 3-D boundary, which for a large-$N$ and a small gauge coupling are approximated by free field theories. These theories are dual to the 4-D domain wall gravitational background. The analogous mapping between the cosmological background and the boundary theory requires an analytic continuation. This takes the form 
\bea
k_i\rightarrow -i k_i \qquad \qquad N \rightarrow  -iN, 
\label{cont}
\eea
in all the momenta of the 3-T correlators defined above, after the redefinition illustrated in (\ref{repper}),
with $N$ denoting the rank of the gauge group in \cite{Bzowski:2011ab}. From now on, in this section, we assume that in all the correlators computed in the previous sections we have done the replacement
(\ref{repper}), followed by the analytic continuation described by (\ref{cont}).
 
 The final formula for the gravitational perturbations then relates the scalar and tensor fluctuations of the metric to the imaginary parts of the redefined correlators\footnote{In the notation of \cite{Bzowski:2011ab} the euclidean momenta are denoted as $\bar{q}_i$, and correspond to our $k_i$ before the analystic continuation. The authors set $\bar q_i= - i q_i$ to define the euclidean momenta of the cosmological 
correlators in 3-D space, with $q_i$ the final momenta in this space}. The derivations of the 
holographic expressions for each type of perturbations are quite involved, but in the case of scalar cosmological perturbations, parameterized by the field $\zeta$, they take a slightly simpler form 
\bea
\label{holo_1}
\langle \zeta(q_1)\z(q_2)\z(q_3)\rangle 
 &=& -\frac{1}{256}\Big(\prod_i \Im [B(\bq_i)]\Big)^{-1}\times
\Im \Big[\langle T(\bq_1)T(\bq_2)T(\bq_3)\>\!\> + 4\sum_i B(\bq_i) \nn\\ 
&-& 2\Big( \langle T(\bq_1)\Upsilon(\bq_2,\bq_3)\rangle +\mathrm{cyclic\,perms.}\Big)\Big].
\eea
Similar formulas are given for the tensor and mixed scalar/tensor perturbations, which can be found in \cite{Bzowski:2011ab}.
In this expression we have omitted an overall factor of $(2\pi^3)$ times a delta function, for momentum conservation.
In \cite{Bzowski:2011ab} the authors use latin indices for the 3-D correlators and introduce the function 
\beq
\Upsilon_{ijkl}(x_1,x_2)\equiv \langle \frac{\delta T^{ij}(x_1)}{\delta g^{k l}(x_2)}\rangle
\label{yuppi}
\eeq
which characterises the contact terms, being proportional to a delta function ($\sim \delta(x_1-x_2)$). 
For the rest their conventions are 
\begin{align}
&T(\bq) = \delta_{ij} T_{ij}(\bq), \qquad
 \qquad \Upsilon(\bq_1,\bq_2) =\delta_{ij}\delta_{kl} \Upsilon_{ijkl}(\bq_1,\bq_2) \,.
\end{align}
 The coefficients  $B(\bq_i)$
are related to 2-point functions of the stress tensor.   $\Upsilon$ stands for
the trace of the $\Upsilon_{ijkl}$ tensor. Eq. (\ref{holo_1}) allows to map the computation of the bispectrum of the scalar perturbations in ordinary cosmology to a computation of correlation functions in simple free field 
theories. In this case the correlators on the right hand side are obtained by adding the scalar, fermion and gauge contributions. They correspond to fully traced 3-T correlators and contact terms whose imaginary parts are generated by the application of the replacements (\ref{cont}) on the diagrammatic results found in the previous sections.

In order to compare our results, expressed in terms of functional derivatives of $\mathcal{S}$ with those of \cite{Bzowski:2011ab} we define
\bea
\label{Upsilon}
\Upsilon^{\mu\nu\alpha\beta}(z,x) = \frac{\delta T^{\mu\nu}(z)}{\delta g_{\alpha \beta}(x)} \bigg|_{g_{\mu\nu} = \delta_{\mu\nu}} = - \frac{1}{2} \delta^{\alpha \beta}
\delta(z-x) T^{\mu\nu}(z) - 2 \, [S]^{\mu\nu\alpha\beta}(z,x)
\eea
which is the analog of $\Upsilon_{ijkl}(z,x)$ defined in Eq. (\ref{yuppi}).
Because the operations of raising and lowering indices do not commute with the metric functional derivatives, 
$\Upsilon ^{\mu\nu\alpha\beta}(z,x)$ and $\Upsilon_{\mu\nu\alpha\beta}(z,x)$ (we use greek indices) are not simply related
by the contractions with metric tensors. 
A careful analysis shows that the relation between the two quantities in the flat space-time limit is
\bea
\Upsilon_{\mu\nu\alpha\beta}(z,x) =  \frac{\delta T_{\mu\nu}(z)}{\delta g^{\alpha \beta}(x)} \bigg|_{g_{\mu\nu} = \delta_{\mu\nu}} &=&
-\frac{1}{2} \delta(z-x) \bigg[ \delta_{\alpha \mu} T_{\beta \nu} + \delta_{\beta \mu} T_{\alpha \nu} 
+ \delta_{\alpha \nu} T_{\beta \mu} + \delta_{\beta \nu} T_{\alpha \mu} \bigg](z) \nn \\
&-& \delta_{\mu \rho} \delta_{\nu \sigma} \delta_{\alpha \gamma} \delta_{\beta \delta}   \Upsilon^{\rho\sigma\gamma\delta}(z,x) \,,
\eea
and taking in account Eq. (\ref{Upsilon}) we can finally map our contact terms with the expressions of \cite{Bzowski:2011ab}
\bea
 \langle \Upsilon_{\mu\nu\alpha\beta}(z,x) T_{\rho\sigma}(y) \rangle &=& 
- \frac{1}{2} \delta(z-x)  \bigg[ \delta_{\alpha \mu} \langle T_{\beta \nu}(z) T_{\rho\sigma}(y)\rangle + \delta_{\beta \mu} \langle T_{\alpha \nu}(z) T_{\rho\sigma}(y)\rangle + \delta_{\alpha \nu} \langle T_{\beta \mu}(z) T_{\rho\sigma}(y)\rangle \nn \\
&+& 
\delta_{\beta \nu} \langle T_{\alpha \mu}(z) T_{\rho\sigma}(y)\rangle 
 -    \delta_{\alpha \beta}  \langle T_{\mu \nu}(z) T_{\rho\sigma}(y) \rangle  \bigg]  + 2 \,  \langle [\mathcal S]_{\mu \nu \alpha 
 \beta }(z,x) T_{\rho \sigma}(y) \rangle.
\eea
This equation is sufficient in order to provide a complete mapping between our results and those of \cite{Bzowski:2011ab} in the $TTT$ case.

\section{Computation of $TTTT$}
\begin{figure}[t]
\label{top3T}
\centering
\includegraphics[scale=0.8]{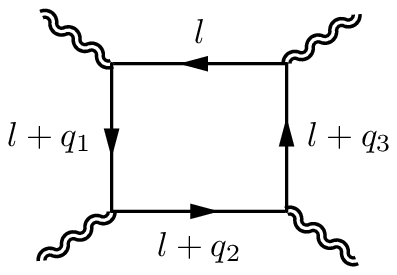}
\hspace{1cm}
\includegraphics[scale=0.8]{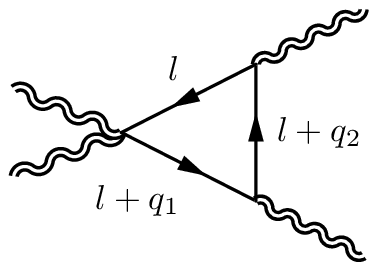}
\hspace{1cm}
\includegraphics[scale=0.8]{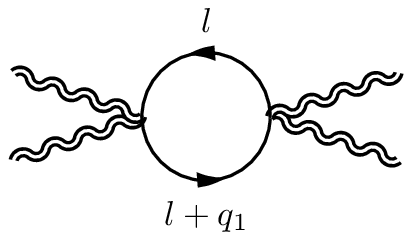}
\hspace{1cm}
\includegraphics[scale=0.8]{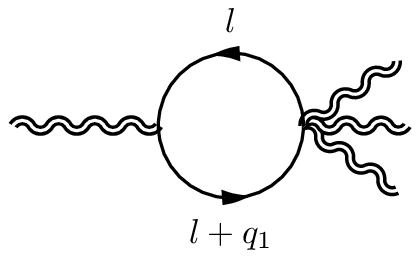}
\caption{Topologies appearing in the expansion of the $4-T$ correlator $\textrm{BoxTop}$, $\textrm{TriTop}$, $\textrm{BubTop}_{22}$ and $\textrm{BubTop}_{13}$. \label{TTTTtop}}
\end{figure}
\begin{figure}[t]
\centering
\includegraphics[scale=0.8]{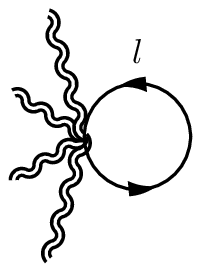}
\caption{Tadpole topology for the TTTT correlator. \label{TTTTtadtop}}
\end{figure}
The evaluation of the correlation function of 4 EMT' s is very involved, due to its tensor structure, which is of rank-8, but it becomes more manageable in the case of the scalar amplitude. This is obtained  by tracing all the indices pairwise, given by $\langle TTTT\rangle$ and all the corresponding contact terms, which are identified from the Feynman expansion of the 
related $<TTTT>$.  
The general structure of the 4-T correlator has been defined in full generality in Eq. (\ref{4PF}) and the scalar component of this relation can be trivially extracted from the same equation. \\ 
We present the results for the minimally coupled scalar, the gauge field, the conformal scalar and the chiral fermion cases. The vertices with one, two and three EMT insertions used in the computation are given in the Appendix \ref{Vertices}. \\
We start analyzing in detail the minimally coupled scalar focusing on the classification of the contributions of different topologies. 
\subsection{$TTTT$ for the minimally coupled scalar case}
The first term on the right hand-side of Eq. (\ref{4PF}), the "ordinary" $\langle TTTT \rangle$ correlator, 
is a four-point function with three box-like contributions which can be obtained, in momentum space, from each other with a suitable re-parameterization of the internal momenta circulating in the loop.  For this reason we compute just one box diagram $\textrm{BoxTop}^{MS}(\vec{q_1},\vec{q_2},\vec{q_3})$ with the internal momenta flowing in the loop as depicted in Fig. \ref{TTTTtop}, obtaining
\bea
\langle T(\ku)T(\kd)T(\kt)T(\kq)\rangle_{MS}  &=&16\left( \textrm{BoxTop}^{MS}(\ku,\ku+\kd,-\kq) + \textrm{BoxTop}^{MS}(\ku,\ku+\kd,-\kt) \right.\nn \\
 &+&\left. \textrm{BoxTop}^{MS}(\ku,\kd+\kt,-\kq)\right) \,.
\eea
The triangle terms in Eq. (\ref{4PF}) are contact terms with the insertion of two external legs on the same vertex. There are six triangle diagrams of this type, characterized by the different couples of attached external momenta $\ku, \kd, \kt, \kq$. Each of these contributions is obtained from the triangle diagram $\textrm{TriTop}^{MS}(\vec{q_1},\vec{q_2})$, illustrated in Fig. \ref{TTTTtop}, with the following assignments of momenta
\bea
&& \textrm{TriTop}^{MS}(\ku , -\kq), \qquad \textrm{TriTop}^{MS}(\ku,-\kt), \qquad  \textrm{TriTop}^{MS}(\ku,\ku+\kd), \\ \nn
&& \textrm{TriTop}^{MS}(\ku + \kd, -\kq), \qquad \textrm{TriTop}^{MS}(\ku+\kt,-\kq) \qquad \textrm{TriTop}^{MS}(\ku+\kq,-\kt) \,.
\eea
Here we provide an example 
\bea
\langle \left[\mathcal S\right]^{\muu \, \mud}_{\muu \, \mud}(\ku,\kd)T(\kt) T(\kq)\rangle_{MS} =  4 \, \textrm{TriTop}^{MS}(\ku + \kd , -\kq) \,.
\eea
There are also two classes of contact terms with bubble topologies, namely $\textrm{BubTop}^{MS}_{22}(\vec{q_1})$ and $\textrm{BubTop}^{MS}_{13}(\vec{q_1})$ depicted in Fig. \ref{TTTTtop}. The former is defined by two vertices with double T insertions, while the latter is characterized by a three external leg insertion on the same point. In the first case there are three contributions coming from the distinct rearrangements of the external momenta into two pairs, given by
\bea
\textrm{BubTop}^{MS}_{22}(\ku + \kd) \qquad \textrm{BubTop}^{MS}_{22}(\ku + \kt) \qquad \textrm{BubTop}^{MS}_{22}(\ku + \kq) \,,
\eea 
while in the other topology class four terms appear which are given by
\bea
\textrm{BubTop}^{MS}_{13}(\ku) \qquad\textrm{BubTop}^{MS}_{13}(\kd) \qquad \textrm{BubTop}^{MS}_{13}(\kt) \qquad  \textrm{BubTop}^{MS}_{13}(\kq) \,.
\eea
From Eq. (\ref{4PF}) we deduce that 
\bea
\langle \left[\mathcal S\right]^{\muu \, \mud}_{\muu \, \mud}(\ku,\kd) \left[\mathcal S\right]^{\mut \, \muq}_{\mut \, \muq}(\kt,\kq) \rangle_{MS} &=& \textrm{BubTop}^{MS}_{22}(\ku + \kd) \,,\\
\langle \left[\mathcal S\right]^{\muu \, \mud \, \mut}_{\muu \, \mud \, \mut}(\ku,\kd,\kt)  T(\kq)\rangle_{MS} &=& -2 \, \textrm{BubTop}^{MS}_{13}(\kq) \,,
\eea
and with similar expressions for the other contributions. Notice that there is also a massless tadpole in Fig. \ref{TTTTtadtop} which is scheme-dependent and can be set to zero.  \\
For bubble and triangular topologies, the results take a simple form 
\bea
\textrm{TriTop}^{MS}(\vec{q_1},\vec{q_2}) &=& \frac{1}{2048 \pi ^3}   \bigg\{ 
- \left( q_1^2 + q_{12}^2 + q_2^2 \right) \bigg[ 
q_1^2 \, \mathcal B_0 \left(q_1^2 \right)
+ q_{12}^2 \, \mathcal B_0 \left( q_{12}^2 \right)
+ q_2^2 \, \mathcal B_0 \left( q_2^2 \right) \bigg]  \nn \\
&+&
2 \, q_1^2 \, q_{12}^2  \, q_2^2 \, \mathcal C_0 \left( q_1^2, q_{12}^2, q_2^2 \right)
\bigg\} \,,  \\
\textrm{BubTop}^{MS}_{22}(\vec{q_1}) &=& \frac{q_1^4}{1024 \pi^3} \mathcal B_0(q_1^2) \,, \\
\textrm{BubTop}^{MS}_{13}(\vec{q_1}) &=& \frac{3 \, q_1^4}{1024 \pi^3} \mathcal B_0(q_1^2) \,,
\eea
where $q^2_{ij} = (q_i -q_j)^2$.\\
The diagram with box topology, $\textrm{BoxTop}^{MS}$, is more involved and it is expanded on a basis of scalar integrals $\mathcal I_i$ with coefficients $C_i^{MS}$. 
The basis is built from 2-, 3- and 4- point massless scalar integrals reported in Appendix \ref{scalarint}.
The elements of this basis are not all independents from each other because $\mathcal D_0$ can be removed using Eq. (\ref{D0toC0}). Nevertheless we show the results in this form in order to simplify their presentation. \\
The basis of scalar integrals is given by
\bea
\mathcal I_1(\vec{q_1},\vec{q_2},\vec{q_3}) &=& \mathcal D_0 \left(q_1^2,q_{12}^2,q_{23}^2,q_3^2,q_2^2,q_{13}^2\right) \,, \nn \\
\mathcal I_2(\vec{q_1},\vec{q_2},\vec{q_3}) &=& \mathcal C_0 \left(q_1^2,q_{12}^2, q_2^2 \right) \,, \nn \\
\mathcal I_3(\vec{q_1},\vec{q_2},\vec{q_3}) &=& \mathcal C_0 \left(q_1^2,q_{13}^2, q_3^2 \right) \,,  \nn \\
\mathcal I_4(\vec{q_1},\vec{q_2},\vec{q_3}) &=& \mathcal C_0 \left(q_2^2,q_{23}^2, q_3^2 \right) \,, \nn \\
\mathcal I_5(\vec{q_1},\vec{q_2},\vec{q_3}) &=& \mathcal C_0 \left(q_{12}^2,q_{23}^2, q_{13}^2 \right) \,, \nn \\
\mathcal I_6(\vec{q_1},\vec{q_2},\vec{q_3}) &=& \mathcal B_0 \left(q_1^2 \right) \,, \nn \\
\mathcal I_7(\vec{q_1},\vec{q_2},\vec{q_3}) &=& \mathcal B_0 \left(q_2^2 \right) \,, \nn \\
\mathcal I_8(\vec{q_1},\vec{q_2},\vec{q_3}) &=& \mathcal B_0 \left(q_3^2 \right) \,, \nn \\
\mathcal I_9(\vec{q_1},\vec{q_2},\vec{q_3}) &=& \mathcal B_0 \left(q_{12}^2 \right) \,, \nn \\
\mathcal I_{10}(\vec{q_1},\vec{q_2},\vec{q_3}) &=& \mathcal B_0 \left(q_{13}^2 \right) \,, \nn \\
\mathcal I_{11}(\vec{q_1},\vec{q_2},\vec{q_3}) &=& \mathcal B_0 \left(q_{23}^2 \right) \,, 
\eea
in terms of which the diagram with the box topology can be expressed as 
\bea
\textrm{BoxTop}^{MS}(\vec{q_1},\vec{q_2},\vec{q_3}) = \sum_{i=1}^{11} C^{MS}_i(\vec{q_1},\vec{q_2},\vec{q_3}) \, \mathcal I_i(\vec{q_1},\vec{q_2},\vec{q_3}).
\label{boxo}
\eea
The first few coefficients are given by
\bea
C^{MS}_1(\vec{q_1},\vec{q_2},\vec{q_3})&=& \frac{q_1^2 \, q_{12}^2 \, q_{23}^2 \, q_3^2}{2048 \pi^3} \,, \nn 
\eea
\bea
C^{MS}_2(\vec{q_1},\vec{q_2},\vec{q_3}) &=& \frac{q_1^2 q_{12}^2}{2048 \pi^3 \, \lambda^2(q_1,q_{12},q_2)} \bigg[  q_3^2 q_1^4-\left(\left(2 q_3^2+2 q_{12}^2-q_{13}^2+q_{23}^2\right) q_2^2+q_{12}^2
   \left(q_3^2+q_{23}^2\right)\right) q_1^2 \nn \\
&+&  \left(q_2^2-q_{12}^2\right) \left(q_2^2 \left(q_3^2-q_{13}^2+q_{23}^2\right)-q_{12}^2 q_{23}^2\right) \bigg] \,, \nn
\eea
\bea
C^{MS}_3(\vec{q_1},\vec{q_2},\vec{q_3}) &=&  \frac{q_1^2 q_3^2}{2048 \pi^3 \, \lambda^2(q_1,q_{13},q_3)} \bigg[ q_{12}^2 q_1^4-\left(\left(q_{12}^2+2
   q_{13}^2+q_{23}^2\right) q_3^2+q_{13}^2 \left(-q_2^2+2 q_{12}^2+q_{23}^2\right)\right) q_1^2 \nn \\
&+& \left(q_3^2-q_{13}^2\right) \left(\left(q_2^2-q_{12}^2-q_{23}^2\right) q_{13}^2+q_3^2 q_{23}^2\right) \bigg] \,, \nn 
\eea
\bea
C^{MS}_4(\vec{q_1},\vec{q_2},\vec{q_3}) &=& \frac{q_3^2 q_{23}^2}{2048 \pi^3 \, \lambda^2(q_2,q_{23},q_3)} \bigg[ q_1^2 q_3^4-\left(\left(2
   q_1^2+q_{12}^2-q_{13}^2\right) q_2^2+\left(q_1^2+2 q_2^2+q_{12}^2\right) q_{23}^2\right) q_3^2 \nn \\
&+&  \left(q_2^2-q_{23}^2\right) \left(q_2^2 \left(q_1^2+q_{12}^2-q_{13}^2\right)-q_{12}^2 q_{23}^2\right) \bigg] \,, \nn \\
C^{MS}_5(\vec{q_1},\vec{q_2},\vec{q_3}) &=& \frac{q_{12}^2 q_{23}^2}{2048 \pi^3 \, \lambda^2(q_{12},q_{23},q_{13})} \bigg[ q_3^2
   q_{23}^4-\left(\left(q_1^2+q_3^2\right) q_{12}^2+\left(q_1^2-q_2^2+2 \left(q_3^2+q_{12}^2\right)\right) q_{13}^2\right) q_{23}^2 \nn \\
&+& \left(q_{12}^2-q_{13}^2\right) \left(q_1^2 q_{12}^2-\left(q_1^2-q_2^2+q_3^2\right) q_{13}^2\right) \bigg] \, ,
\eea
the remaining ones, having lengthier forms, have been collected in Appendix \ref{MSres}.

We remark, if not obvious, that being the $TTTT$ correlator in D=3 dimensions finite and hence (trace) anomaly free, the operation of tracing the indices of an energy-momentum tensor can be performed before or after the evaluation of the integrals appearing in the loops, with no distinction. In D=4 the two procedures are inequivalent, differing by the anomalous term. 

For this reason we have computed the $TTTT$ correlator in two distinct ways, obviously with the same result. In the first case we have traced all the four pairs of indices before computing the loop integrals. In this way we obtain directly the $TTTT$ correlator. In the second case, which is much more involved, we have calculated the $T^{\mu \nu}TTT$ correlation function with a pair of indices not contracted, we have evaluated the tensor integrals and then we have contracted the result with $\delta_{\mu\nu}$. This intermediate step is useful in order to test our computation with the diffeomorphism Ward identity given in Eq. (\ref{WI4PFMomentumFlat}). 

\subsection{$TTTT$ for the gauge field case}

As discussed in the previous section also in the case of an abelian gauge field the scalar component of the 4-T correlator can be decomposed in the sum of three box diagrams as
\bea
\langle T(\ku) T(\kd)T(\kt)T(\kq)\rangle_{GF}  &=& 16 \bigg( \textrm{BoxTop}^{GF}(\ku, \ku+\kd,-\kq) + \textrm{BoxTop}^{GF}(\ku,\ku+\kd,-\kt) \nn \\
 &+& \textrm{BoxTop}^{GF}(\ku,\ku+\kt,-\kq) \bigg) \,,
\eea
where the box diagram contributions can be written in terms of a minimal scalar box term (MS) and of extra terms as 
\bea
\textrm{BoxTop}^{GF}(\vec{q_1},\vec{q_2},\vec{q_3}) &=& \textrm{BoxTop}^{MS}(\vec{q_1},\vec{q_2},\vec{q_3})  + 
\frac{1}{2048 \pi^3}\bigg\{
\left(2 q_1^2-q_2^2-q_3^2-q_{12}^2-q_{13}^2\right) q_1^2
   \, \mathcal B_0\left(q_1^2\right) \nn \\
&+&  \left(2
   q_{23}^2 - q_2^2 - q_3^2 - q_{12}^2 - q_{13}^2\right) q_{23}^2
   \, \mathcal B_0\left(q_{23}^2\right)
+  \left(2 q_2^2 -q_1^2-q_3^2-q_{12}^2-q_{23}^2\right) q_2^2
 \,  \mathcal B_0\left(q_2^2 \right)  \nn \\
&+&  \left(2 q_3^2 -q_1^2-q_2^2-q_{13}^2-q_{23}^2\right) q_3^2
  \, \mathcal B_0 \left(q_3^2\right)
+  \left(2 q_{12}^2 -q_1^2-q_2^2-q_{13}^2-q_{23}^2\right) q_{12}^2
  \, \mathcal B_0 \left(q_{12}^2 \right) \nn \\
&+&  \left(2 q_{13}^2 -q_1^2-q_3^2-q_{12}^2-q_{23}^2\right) q_{13}^2
 \, \mathcal B_0 \left(q_{13}^2\right) 
+ 2 \, q_2^2 q_{12}^2 q_1^2
  \, \mathcal C_0 \left(q_1^2,q_{12}^2,q_2^2 \right) \nn \\
&+& 2 q_3^2 q_{13}^2 q_1^2
  \,  \mathcal C_0 \left(q_1^2,q_{13}^2,q_3^2 \right)
+ 2 q_2^2 q_3^2 q_{23}^2
  \, \mathcal C_0 \left(q_2^2,q_{23}^2,q_3^2 \right)
+ 2 q_{12}^2 q_{13}^2 q_{23}^2
  \, \mathcal C_0 \left( q_{12}^2,q_{23}^2,q_{13}^2 \right)  \bigg\} \,. \nn \\
\eea
The reconstruction of the  
$< T(x_1)T(x_2)T(x_3)T(x_4)> $ amplitude can be obtained using the expression above plus the contributions of triangle and bubble topology, corresponding to the relative contact terms. In the case of the gauge field these are 
\bea
 \langle \left[\mathcal S\right]^{\muu \, \mud}_{\muu \, \mud}(\ku,\kd)T(\kt) T(\kq)\rangle_{GF} &=&  4 \, \textrm{TriTop}^{GF}(\ku + \kd , -\kq) \,, \\
 \langle \left[\mathcal S\right]^{\muu \, \mud}_{\muu \, \mud}(\ku,\kd) \left[\mathcal S\right]^{\mut \, \muq}_{\mut \, \muq}(\kt,\kq) \rangle_{GF} &=& \textrm{BubTop}^{GF}_{22}(\ku + \kd) \,,\\
 \langle \left[\mathcal S\right]^{\muu \, \mud \, \mut}_{\muu \, \mud \, \mut}(\ku,\kd,\kt)T(\kq)\rangle_{GF} &=& -2 \, \textrm{BubTop}^{GF}_{13}(\kq) \,,
\eea
and with similar expressions for the other contributions, obtained by a suitable reparameterization of the internal momenta. In this case the discussion is identical as in the minimally coupled scalar. The explicit results for these three topologies are given by
\bea
\textrm{TriTop}^{GF}(\vec{q_1},\vec{q_2}) &=& \frac{1}{2048 \pi ^3}   \bigg\{ 6 \, q_1^2 \, q_{12}^2  \, q_2^2 \, \mathcal C_0 \left( q_1^2, q_{12}^2, q_2^2 \right)
- 3 \, q_1^2 (- 3 q_1^2 + q_{12}^2 + q_2^2 ) \, \mathcal B_0(q_1^2) \nn \\
&-& 3 \, q_{12}^2 (q_1^2 - 3 q_{12}^2 + q_2^2 ) \, \mathcal B_0(q_{12}^2)  
- 3 \, q_2^2 ( q_1^2 + q_{12}^2 - 3 q_2^2 ) \, \mathcal B_0(q_2^2)
\bigg\} \,,   \\
\textrm{BubTop}^{GF}_{22}(\vec{q_1}) &=& \frac{9 \, q_1^4}{1024 \pi^3} \mathcal B_0(q_1^2) \,, \\
\textrm{BubTop}^{GF}_{13}(\vec{q_1}) &=& \frac{15 \, q_1^4}{1024 \pi^3} \mathcal B_0(q_1^2) \,.
\eea

\subsection{$TTTT$ for the conformally coupled scalar case}

As for the gauge fields, also for the conformally coupled scalar the totally traced component of the 4-T correlator is given by
\bea
\langle T(\ku) T(\kd) T(\kt)T(\kq)\rangle_{CS}  
&=& 
16 \bigg( \textrm{BoxTop}^{CS}(\ku, \ku+\kd,-\kq) + \textrm{BoxTop}^{CS}(\ku,\ku+\kd,-\kt) 
\nn \\
&+& 
\textrm{BoxTop}^{CS}(\ku,\ku+\kt,-\kq) \bigg) \,,
\eea
where the box diagram contributions can be written as
\bea
\textrm{BoxTop}^{CS}(\vec{q_1},\vec{q_2},\vec{q_3})
&=& 
\frac{1}{4096 \,\pi^3}\, \bigg\{
\mathcal B_0(q_{13}^2)\, \left(3 (\vec{q_1} \cdot \vec{q_3} )^2 + \vec{q_1} \cdot \vec{q_3} (2 q_2^2 - 3 \vec{q_2} \cdot \vec{q_3}) + q_1^2\, ( \vec{q_2} \cdot \vec{q_3} - q_3^2) 
\right. 
\nn \\
&+&
\left. 
\vec{q_1} \cdot \vec{q_2} (-3 \vec{q_1} \cdot \vec{q_3}  + q_3^2) \right)
- \mathcal B_0(q_2^2)\, \left(\vec{q_1} \cdot \vec{q_3} \, q_2^2 + 2\, q_1^2 (\vec{q_2} \cdot \vec{q_3} - q_3^2) 
\right.
\nn \\
&+&
\left. \vec{q_1} \cdot \vec{q_2} \, (-3 \, \vec{q_2} \cdot \vec{q_3} + 2 q_3^2)\right) \bigg\}\, . 
\eea
Concerning the contact terms, the triangle and bubble topology contributions are  given by
\bea
\langle \left[\mathcal S\right]^{\muu \, \mud}_{\muu \, \mud}(\ku,\kd)T(\kt) T(\kq)\rangle_{CS} &=& 
\frac{1}{256 \, \pi^3} \vec{k_1} \cdot \vec{k_2} \, \vec{k_3} \cdot \vec{k_4} \, \mathcal B_0\left( (\vec{k_1} + \vec{k_2} )^2 \right)
\,, \\
\langle \left[\mathcal S\right]^{\muu \, \mud}_{\muu \, \mud}(\ku,\kd) \left[\mathcal S\right]^{\mut \, \muq}_{\mut \, \muq}(\kt,\kq) 
\rangle_{CS} 
&=& \frac{1}{1024 \, \pi^3} \vec{k_1} \cdot \vec{k_2} \,  \vec{k_3} \cdot \vec{k_4} \, \mathcal B_0\left( (\vec{k_1} + \vec{k_2} )^2 \right)   \,,\\
\langle \left[\mathcal S\right]^{\muu \, \mud \, \mut}_{\muu \, \mud \, \mut}(\ku,\kd,\kt)T(\kq)\rangle_{CS} &=& 0 \,.
\eea

\subsection{$TTTT$ for the chiral fermion case}

In the case of the chiral fermion field the ordinary EMT's correlation function is zero
\bea
\langle T(\ku) T(\kd) T(\kt)T(\kq)\rangle_{CF}  
&=& 
16 \bigg( \textrm{BoxTop}^{CF}(\ku, \ku+\kd,-\kq) + \textrm{BoxTop}^{CF}(\ku,\ku+\kd,-\kt) 
\nn \\
&+& 
\textrm{BoxTop}^{CF}(\ku,\ku+\kt,-\kq) \bigg) = 0 \,,
\eea
despite the fact that the single box contribution is given by
\beqa
\textrm{BoxTop}^{CF}(\vec{q_1},\vec{q_2},\vec{q_3}) 
&=& 
\frac{1}{128 \,\pi^3}\, \bigg\{ 
B_0(q_2^2)\, (- q_2^2 \, \vec{q_1} \cdot \vec{q_3}   + \vec{q_1} \cdot \vec{q_2} \, \vec{q_2} \cdot \vec{q_3} ) 
\nn \\ 
&+& 
B_0(q_{31}^2)\, \left(- \vec{q_1} \cdot \vec{q_3} ( \vec{q_1} \cdot \vec{q_2} - \vec{q_1}\cdot \vec{q_3} + \vec{q_2} \cdot \vec{q_3}) 
+ q_1^2  (\vec{q_2} \cdot \vec{q_3} - q_3^2) + q_3^2 \, \vec{q_1} \cdot \vec{q_2} \right)
\bigg\} \,.
\eeqa
All the other topologies are identically zero,
\bea
\langle \left[\mathcal S\right]^{\muu \, \mud}_{\muu \, \mud}(\ku,\kd)T(\kt) T(\kq)\rangle_{CF} &=& 0 \,, \\
\langle \left[\mathcal S\right]^{\muu \, \mud}_{\muu \, \mud}(\ku,\kd) \left[\mathcal S\right]^{\mut \, \muq}_{\mut \, \muq}(\kt,\kq)  \rangle_{CF}  &=& 0 \,,\\
\langle \left[\mathcal S\right]^{\muu \, \mud \, \mut}_{\muu \, \mud \, \mut}(\ku,\kd,\kt)T(\kq)\rangle_{CF} &=& 0  \,,
\eea
with all the similar contributions obtained by exchanging the respective momenta.

\subsection{Relations between contact terms in the 4-T case} 
As we have mentioned in the introduction, the extension of the holographic formula for scalar and tensor perturbations remains to be worked out. This is expected to require a lengthy but straightforward extension of the methods developed for the analysis of the bispectrum. For this reason here we reformulate the structure of the contact terms, which are expected to be part of this extension, in a form similar to those presented in the previous section.  We recall that in our notations the contact terms are given as in Eq. (\ref{4PF}). 

For example, extending the previous notations, the contact term with the triangle topology in the two formulations are related as
\bea \label{U1TT}
\langle \Upsilon_{\mu\nu\alpha\beta}(z,x) T_{\rho\sigma}(y) T_{\lambda\tau}(t) \rangle 
&=& 
- \frac{1}{2} \delta(z-x)  \bigg[ \delta_{\alpha \mu} \langle T_{\beta \nu}(z) T_{\rho\sigma}(y)T_{\lambda\tau}(t)\rangle + 
\delta_{\beta \mu} \langle T_{\alpha \nu}(z) T_{\rho\sigma}(y)T_{\lambda\tau}(t)\rangle \nn \\
&+& 
\delta_{\alpha \nu} \langle T_{\beta \mu}(z) T_{\rho\sigma}(y)T_{\lambda\tau}(t)\rangle
+ \delta_{\beta \nu} \langle T_{\alpha \mu}(z) T_{\rho\sigma}(y)T_{\lambda\tau}(t)\rangle 
- \delta_{\alpha \beta}  \langle T_{\mu \nu}(z) T_{\rho\sigma}(y)T_{\lambda\tau}(t) \rangle  \bigg]  \nn \\
&+& 
2 \,  \langle [\mathcal S]_{ \mu \nu \alpha \beta}(z,x) T_{\rho \sigma}(y) T_{\lambda\tau}(t) \rangle.
\eea
Similar relations hold for those contact terms of bubble topology  
\bea \label{U1U1}
\langle \Upsilon_{\mu\nu\alpha\beta}(z,x) \Upsilon_{\rho\sigma\lambda\tau}(y,t) \rangle
&=&
\frac{1}{4}\, \delta(z-x) \delta(y-t)\, 
\bigg[
\bigg(
  \delta_{\alpha\mu}\delta_{\lambda\rho} \langle T_{\beta\nu}(z) T_{\tau\sigma}(x) \rangle 
+ \delta_{\alpha\mu}\delta_{\lambda\sigma} \langle T_{\beta\nu}(z) T_{\tau\rho}(x) \rangle 
\nn \\
&& \hspace{-45mm} 
+\,  \delta_{\alpha\mu}\delta_{\tau\rho}    \langle T_{\beta\nu}(z) T_{\lambda\sigma}(x)  \rangle
+    \delta_{\alpha\mu} \delta_{\tau\sigma} \langle T_{\beta\nu}(z) T_{\lambda\rho}(x) \rangle
+ ( \mu \leftrightarrow \nu )
\bigg)
+ (\alpha \leftrightarrow \beta)
\bigg]
\nn \\
&& \hspace{-45mm}
-\bigg\{ 
\delta(z-x)\, 
\bigg[
\frac{1}{4}\, \delta_{\lambda\tau}\, \delta(y-t)\, 
\bigg( 
  \delta_{\alpha\mu}\langle T_{\beta\nu}(z) T_{\rho\sigma}(y) \rangle 
+ \delta_{\alpha\nu}\langle T_{\beta\mu}(z) T_{\rho\sigma}(y) \rangle
+ \left( \alpha \leftrightarrow \beta \right) \bigg)
\nn \\
&& \hspace{-45mm} 
+ \bigg(
  \delta_{\alpha\mu} \langle T_{\beta\nu}(z) \left[\mathcal S\right]_{\rho\sigma\lambda\tau}(y,t) \rangle
+ \delta_{\alpha\nu} \langle T_{\beta\mu}(z) \left[\mathcal S\right]_{\rho\sigma\lambda\tau}(y,t) \rangle
+ \left(\alpha \leftrightarrow \beta \right) \bigg) \bigg] 
+\, \left(\mu,\nu,z,\alpha,\beta,x\right) \leftrightarrow \left(\rho,\sigma,y,\lambda,\tau,t\right) 
\bigg\}
\nn \\
&& \hspace{-45mm} 
+ \bigg( 
\frac{1}{4}\, \delta_{\alpha\beta}\, \delta_{\lambda\tau}\, \delta(z-x)\, \delta(y-t)\, \langle T_{\mu\nu}(z)T_{\rho\sigma}(y)\rangle
+ \, \delta_{\alpha\beta}\, \delta(z-x)\, 
\langle T_{\mu\nu}(z)\left[\mathcal S\right]_{\rho\sigma\lambda\tau}(y,t)\rangle
\nn \\
&& \hspace{-45mm}
+\, \delta_{\lambda\tau}\, \delta(y-t)\, \langle T_{\rho\sigma}(y) \left[\mathcal S\right]_{\mu\nu\alpha\beta}(z,x)\rangle
+ 4\, \langle \left[\mathcal S\right]_{\mu\nu\alpha\beta}(z,x) \left[\mathcal S\right]_{\rho\sigma\lambda\tau}(y,t)\rangle 
\bigg) \, .
\eea
Finally, we expect, in a possible generalization of the holographic formula for scalar and tensor perturbations at the trispectrum level, double functional derivatives of the EMT with respect to the metric 
\beq
\Upsilon_{\mu\nu\alpha\beta\rho\sigma}(z,x,y) = 
\frac{\delta^2 T_{\mu\nu}(z)}{\delta g^{\rho\sigma}(y)\delta g^{\alpha\beta}(x)}\bigg|_{g_{\mu\nu} = \delta_{\mu\nu}}\, .
\eeq
After some work, the expression of this object in terms of multiple functional derivatives of the action and of EMT's is found to be
\bea\label{U2}
\Upsilon_{\mu\nu\alpha\beta\rho\sigma}(z,x,y)
&=&
\frac{1}{2}\,
\delta(z-x)\delta(z-y)\, 
\bigg[
 \left[ g_{\mu\beta}g_{\alpha\kappa}g_{\nu\lambda} + g_{\mu\alpha}g_{\beta\kappa}g_{\nu\lambda}
+ (\mu \leftrightarrow \nu) - g_{\alpha\beta}g_{\mu\lambda}g_{\nu\kappa} \right]_{\rho\sigma}
\nn \\
&-&
\frac{1}{2} \delta_{\rho\sigma} \left( \delta_{\mu\beta}\delta_{\alpha\kappa}\delta_{\nu\lambda} 
+ \delta_{\mu\alpha}\delta_{\beta\kappa}\delta_{\nu\lambda}+ (\mu \leftrightarrow \nu) 
- \delta_{\alpha\beta}\delta_{\mu\lambda}\delta_{\nu\kappa} \right) \bigg]\, T_{\lambda\kappa}(z)
\nn \\
&-&
\delta(z-x)\, \left[ \delta_{\mu\beta}\delta_{\alpha\kappa}\delta_{\nu\lambda} 
+ \delta_{\mu\alpha}\delta_{\beta\kappa}\delta_{\nu\lambda}+ (\mu \leftrightarrow \nu) 
- \delta_{\alpha\beta}\delta_{\mu\lambda}\delta_{\nu\kappa} \right]
\,\left[\mathcal S\right]_{\lambda\kappa\rho\sigma}(z,y)
\nn \\
&+&
\delta(z-y)\, \left[\delta_{\rho\sigma}\, \left[\mathcal S\right]_{\mu\nu\alpha\beta}(z,x)
- 2\, \left[g_{a\alpha}g_{b\beta}g_{e\mu}g_{f\nu}\right]_{\rho\sigma}\, \left[\mathcal S\right]_{efab}(z,x)\right]
\nn \\
&-& 
2\,\left[\mathcal S\right]_{\mu\nu\alpha\beta\rho\sigma}(z,x,y) \, ,
\eea
where we refer to appendix \ref{Vertices} for a list of the derivatives of metric tensors.\\
We conclude by presenting, along the same lines, the expression of the last contact term which will be present in the holographic extension. 
This is related to our contact terms by the formula
\bea \label{U2T}
\langle \Upsilon_{\mu\nu\alpha\beta\rho\sigma}(z,x,y)T_{\tau\omega}(t) \rangle 
&=&
\frac{1}{2}\,
\delta(z-x)\delta(z-y)\, 
\bigg[
 \left[ g_{\mu\beta}g_{\alpha\kappa}g_{\nu\lambda} + g_{\mu\alpha}g_{\beta\kappa}g_{\nu\lambda}
+ (\mu \leftrightarrow \nu) - g_{\alpha\beta}g_{\mu\lambda}g_{\nu\kappa} \right]_{\rho\sigma}
\nn \\
&-&
\frac{1}{2} \delta_{\rho\sigma} \left( \delta_{\mu\beta}\delta_{\alpha\kappa}\delta_{\nu\lambda} 
+ \delta_{\mu\alpha}\delta_{\beta\kappa}\delta_{\nu\lambda}+ (\mu \leftrightarrow \nu) 
- \delta_{\alpha\beta}\delta_{\mu\lambda}\delta_{\nu\kappa} \right) \bigg]\, \langle T_{\lambda\kappa}(z)T_{\tau\omega}(t)\rangle
\nn \\
&-&
\delta(z-x)\, \left[ \delta_{\mu\beta}\delta_{\alpha\kappa}\delta_{\nu\lambda} 
+ \delta_{\mu\alpha}\delta_{\beta\kappa}\delta_{\nu\lambda}+ (\mu \leftrightarrow \nu) 
- \delta_{\alpha\beta}\delta_{\mu\lambda}\delta_{\nu\kappa} \right]
\,\langle \left[\mathcal S\right]_{\lambda\kappa\rho\sigma}(z,y)T_{\tau\omega}(t)\rangle
\nn \\
&+&
\delta(z-y)\, \left[\delta_{\rho\sigma}\, \langle \left[\mathcal S\right]_{\mu\nu\alpha\beta}(z,x) T_{\tau\omega}(t) \rangle
- 2\, \left[g_{a\alpha}g_{b\beta}g_{e\mu}g_{f\nu}\right]_{\rho\sigma}\, 
\langle \left[\mathcal S\right]_{efab}(z,x) T_{\tau\omega}(t) \rangle \right]
\nn \\
&-& 
2\, \langle \left[\mathcal S\right]_{\mu\nu\alpha\beta\rho\sigma}(z,x,y) T_{\tau\omega}(t) \rangle.
\eea
%
%


\section{Conclusions and Perspectives}
The study of holographic cosmological models is probably at its beginning and there is little doubt that the interest in these models will be growing in the near future. In these formulations, the metric perturbations of a cosmological inflationary phase characterized by strong gravity can be expressed in terms of correlation functions involving stress energy tensors 
in simple 3-D field theories. We have presented an independent derivation of all the amplitudes which are part of the 3-T correlators and extended the analysis to the fully traced component of the 4-T one. The analysis is rather involved and is based on an extension of the approach developed in \cite{Coriano:2012wp}, which dealt with the 3-T case in D=4. In D=3, the absence of anomalies simplifies considerably the treatment, but the perturbative expression of the 4-T amplitude carries the same level of difficulty of the 4-D case. The extension of our approach to a discussion of the full 4-T case, with the derivation of all the amplitudes, is, at the moment, hampered by the remarkable difficulties present in the computation of all the tensor reductions of a rank-8 correlator. We hope to discuss these issues in future work.

\centerline{\bf \large Acknowledgements} 
We thank Roberta Armillis for discussions. C.C. thanks the members of the theory division at EPFL Lausanne for hospitality and in particular Roman Scoccimarro for discussions.  

\begin{appendix}
\section{Scalar integrals}
\label{scalarint}
We give the expressions of the two, three and four point scalar integrals with internal masses set to zero in $D=3$. 
They are  defined as
\beqa
 \mathcal B_0(q_1^2) &=& \int d^3 l \frac{1}{l^2 \, (l+q_1)^2}  = \frac{\pi^3}{q_1}  \,, \nn \\
 \mathcal C_0(q_1^2, q_{12}^2, q_2^2) &=& \int d^3 l \frac{1}{l^2 \, (l+q_1)^2 \, (l+q_2)^2} = \frac{\pi^3}{q_1 \, q_{12} \, q_2}\,, \nn \\
 \mathcal D_0 \left(q_1^2,q_{12}^2,q_{23}^2,q_3^2,q_2^2,q_{13}^2\right) &=& \int d^3 l \frac{1}{l^2 \, (l+q_1)^2 \, (l+q_2)^2 \, (l+q_3)^2}
\eeqa
where $q_{ij}^2 = (q_i-q_j)^2$. \\
The box integral in $D=3$ in not independent from the 2- and 3- scalar point functions, indeed it is possible to show the following relation 
\bea
&& \mathcal D_0 \left(q_1^2,q_{12}^2,q_{23}^2,q_3^2,q_2^2,q_{13}^2\right) =
\frac{1}{q_1^4 q_{23}^4-2 q_1^2 \left(q_3^2 q_{12}^2+q_2^2 q_{13}^2\right) q_{23}^2+\left(q_3^2
   q_{12}^2-q_2^2 q_{13}^2\right){}^2} \times  \nn \\
&&
\bigg\{
 \bigg(q_2^2 q_{13}^4 - \left(\left(q_2^2+q_3^2-2 q_{23}^2\right) q_{12}^2+q_2^2 q_{23}^2\right) q_{13}^2 - q_1^2 q_{23}^2
   \left(q_{12}^2+q_{13}^2-q_{23}^2\right) - q_3^2 q_{12}^2 \left(q_{23}^2-q_{12}^2\right)\bigg)
   \mathcal C_0 \left(q_{12}^2,q_{23}^2,q_{13}^2 \right)       \nn \\
&&
+\bigg(q_2^2 q_{13}^4-\left(\left(q_2^2-2 q_3^2+q_{23}^2\right) q_1^2+q_3^2
   \left(q_2^2+q_{12}^2\right)\right) q_{13}^2+\left(q_1^2-q_3^2\right) \left(q_1^2 q_{23}^2-q_3^2 q_{12}^2\right)\bigg)
   \mathcal C_0 \left(q_1^2,q_3^2,q_{13}^2 \right) \nn \\
&&
+\bigg(\left(2 q_2^2 q_{23}^2-q_{12}^2 \left(q_2^2-q_3^2+q_{23}^2\right)\right) q_3^2+q_1^2
   q_{23}^2 \left(-q_2^2-q_3^2+q_{23}^2\right)-q_2^2 q_{13}^2 \left(-q_2^2+q_3^2+q_{23}^2\right)\bigg)
   \mathcal C_0 \left(q_{23}^2,q_3^2,q_2^2\right) \nn \\
&&
+\bigg(q_{23}^2 q_1^4-\left(\left(q_{13}^2+q_{23}^2\right) q_2^2+q_{12}^2 \left(-2
   q_2^2+q_3^2+q_{23}^2\right)\right) q_1^2+\left(q_{12}^2-q_2^2\right) \left(q_3^2 q_{12}^2-q_2^2 q_{13}^2\right)\bigg)
   \mathcal C_0 \left(q_1^2,q_{12}^2,q_2^2 \right)
\bigg\}.
\label{D0toC0}
\eea

\section{$TTTT$ for the minimal scalar case}
\label{MSres}
We give here the expressions of the remaining coefficients $C^{MS}_i$ of Eq. (\ref{boxo}). They are given by 
\bea
C^{MS}_6(\vec{q_1},\vec{q_2},\vec{q_3}) &=& \frac{1}{2048\pi ^3} \bigg\{ \frac{4}{\lambda^2(q_1,q_{12},q_2)} \bigg[  -q_1^4 \left(\vec{q_1} \cdot \vec{q_2}
   \left(q_2^2+\vec{q_1} \cdot \vec{q_3}+2 \vec{q_2} \cdot \vec{q_3}\right)-q_2^2
   \left(q_2^2+\vec{q_1} \cdot \vec{q_3}\right)+\vec{q_1} \cdot \vec{q_2}^2\right) \nn \\
&+& q_1^2
   \vec{q_1} \cdot \vec{q_2} \left(\vec{q_1} \cdot \vec{q_2}
   \left(\vec{q_1} \cdot \vec{q_3}+\vec{q_2} \cdot \vec{q_3} -q_2^2 \right)-2 q_2^2
   \vec{q_1} \cdot \vec{q_3}+\vec{q_1} \cdot \vec{q_2}^2\right)+q_1^6
   \left(q_2^2+\vec{q_2} \cdot \vec{q_3}\right)+\vec{q_1} \cdot \vec{q_2}^3
   \vec{q_1} \cdot \vec{q_3} \bigg] \nn \\
&+& \frac{4 \vec{q_1} \cdot \vec{q_3}^2}{\lambda^2(q_1,q_{13},q_3)} \left(q_1^2 \vec{q_2} \cdot \vec{q_3}-\vec{q_1} \cdot \vec{q_2}
   \vec{q_1} \cdot \vec{q_3}\right)+q_1^2 q_3^2 \bigg\} \,, \nn
\eea
\bea
C^{MS}_7(\vec{q_1},\vec{q_2},\vec{q_3}) &=& 
\frac{1}{512 \pi ^3
   \lambda^2(q_1,q_{12},q_2)}
 \bigg\{ q_1^4 q_2^2 \vec{q_1} \cdot \vec{q_3}-\vec{q_2} \cdot \vec{q_3} \left(q_1^2
   \left(q_2^4-3 \vec{q_1} \cdot \vec{q_2}^2\right)-q_2^2
   \vec{q_1} \cdot \vec{q_2}^2+q_1^4
   \left(q_2^2+\vec{q_1} \cdot \vec{q_2}\right)\right) \bigg\} \nn \\
&-& \frac{q_3^4}{512 \pi ^3 \lambda^2(q_2,q_{23},q_3)} 
   \left(\vec{q_1} \cdot \vec{q_2} \left(q_2^2+\vec{q_2} \cdot \vec{q_3}\right)-q_2^2
   \vec{q_1} \cdot \vec{q_3}\right)
+\frac{\left(q_1^2+q_2^2\right)
   \left(q_2^2+q_3^2\right)}{2048 \pi ^3}   \nn \\
&+&  \frac{1}{256 \pi ^3 \lambda^2(q_1,q_{12},q_2) \lambda^2(q_2,q_{23},q_3)} 
\bigg\{3 \vec{q_1} \cdot \vec{q_2}^3 \vec{q_2} \cdot \vec{q_3}^2
   \left(\vec{q_2} \cdot \vec{q_3}-2 q_3^2\right) \nn \\
&-& q_2^2 \vec{q_1} \cdot \vec{q_2}
   \vec{q_2} \cdot \vec{q_3} \left(\vec{q_1} \cdot \vec{q_2}^2 \left(q_3^2+2
   \vec{q_2} \cdot \vec{q_3}\right)+\vec{q_1} \cdot \vec{q_2} \vec{q_1} \cdot \vec{q_3}
   \left(\vec{q_2} \cdot \vec{q_3}-4 q_3^2\right)+q_1^2 \vec{q_2} \cdot \vec{q_3}
   (-6 q_3^2   \right. \nn \\
&-& \left. 4 \vec{q_1} \cdot \vec{q_3}+\vec{q_2} \cdot \vec{q_3})\right)-q_2^4 \left(q_3^2
   \left(\vec{q_1} \cdot \vec{q_3} \left(4 q_1^2
   (\vec{q_1} \cdot \vec{q_2}+\vec{q_2} \cdot \vec{q_3})+\vec{q_1} \cdot \vec{q_2}^2\right)+q_1^2
   \vec{q_1} \cdot \vec{q_2} \vec{q_2} \cdot \vec{q_3}-2 \vec{q_1} \cdot \vec{q_2}^3\right)  \right. \nn \\
&+& \left. q_1^2
   \vec{q_2} \cdot \vec{q_3}^2 (\vec{q_1} \cdot \vec{q_3}-2 \vec{q_1} \cdot \vec{q_2})\right)+q_1^2
   q_3^2 q_2^6 (3 \vec{q_1} \cdot \vec{q_3}-2 \vec{q_1} \cdot \vec{q_2}) \bigg\} \,, \nn
\eea
\bea
C^{MS}_8(\vec{q_1},\vec{q_2},\vec{q_3}) &=& \frac{q_3^2}{128 \pi ^3
   \lambda^2(q_1,q_{13},q_3)\lambda^2(q_2,q_{23},q_3)} \bigg\{ \vec{q_1} \cdot \vec{q_2} \vec{q_1} \cdot \vec{q_3}^2 \left(q_3^2
   \left(q_2^2+2 \vec{q_2} \cdot \vec{q_3}\right)-q_3^4-2
   \vec{q_2} \cdot \vec{q_3}^2\right) \nn \\
&+& q_1^2 q_3^2 \vec{q_1} \cdot \vec{q_2}
   \left(\vec{q_2} \cdot \vec{q_3}-q_3^2\right){}^2+\vec{q_1} \cdot \vec{q_3}^3
   \left(q_2^2-\vec{q_2} \cdot \vec{q_3}\right)
   \left(\vec{q_2} \cdot \vec{q_3}-q_3^2\right)+q_1^2 \vec{q_1} \cdot \vec{q_3}
   \left(q_3^4 \left(q_2^2-\vec{q_2} \cdot \vec{q_3}\right) \right. \nn \\
&+& \left. q_3^2
   \vec{q_2} \cdot \vec{q_3} \left(\vec{q_2} \cdot \vec{q_3}-2
   q_2^2\right)+\vec{q_2} \cdot \vec{q_3}^3\right)  \bigg\}
   +\frac{q_3^2}{2048 \pi ^3}
   \left(q_1^2+q_2^2+q_3^2-\vec{q_2} \cdot \vec{q_3}\right) \,, \nn
\eea
\bea
C^{MS}_9(\vec{q_1},\vec{q_2},\vec{q_3}) &=& \frac{q_{12}^2}{4096 \pi^3 \lambda^2(q_1,q_{12},q_2) \lambda^2(q_{12},q_{23},q_{13}) }
  \bigg\{ \left(\left(q_{12}^2-q_{13}^2\right)^2 -q_{23}^4-4 q_{12}^2
   q_{23}^2 \right)
   q_1^6+\left(-q_{23}^6+\left(3 q_2^2+3 q_{13}^2 \right. \right. \nn \\
&+& \left. \left. 2
   \left(q_3^2+q_{12}^2\right)\right) q_{23}^4+\left(9
   q_{12}^4+8 q_2^2 q_{12}^2-3 q_{13}^4-2 \left(2
   q_3^2+q_{12}^2\right) q_{13}^2\right)
   q_{23}^2-\left(q_{12}^2-q_{13}^2\right){}^2 \left(3 q_2^2-2
   q_3^2 \right. \right. \nn \\
&+& \left. \left. 2 q_{12}^2-q_{13}^2\right)\right) q_1^4+\left(-4
   q_{12}^2 q_{23}^6+\left(-3 q_2^4-2 \left(2
   q_3^2+q_{12}^2\right) q_2^2+9 q_{12}^4+8 q_{12}^2
   q_{13}^2\right) q_{23}^4-4 \left(2 q_{12}^6  \right. \right. \nn \\
&+& \left. \left. 2 \left(q_3^2-2
   q_2^2\right) q_{12}^4+q_2^4 q_{12}^2+q_{13}^4 q_{12}^2-2
   \left(2 q_{12}^4+q_2^2 \left(q_3^2+2 q_{12}^2\right)\right)
   q_{13}^2\right) q_{23}^2-\left(-3 q_2^4+\left(4 q_3^2+6
   q_{12}^2\right) q_2^2 \right. \right. \nn \\
&+& \left. \left. q_{12}^4\right)
   \left(q_{12}^2-q_{13}^2\right){}^2\right)
   q_1^2+\left(q_2^2-q_{12}^2\right){}^2
   \left(q_{23}^6+\left(q_2^2+2 q_3^2-2 q_{12}^2-3
   q_{13}^2\right) q_{23}^4-\left(q_{12}^4-3 q_{13}^4 \right. \right.  \nn \\
&+& \left. \left. 2
   \left(2 q_3^2+3 q_{12}^2\right) q_{13}^2\right)
   q_{23}^2-\left(q_{12}^2-q_{13}^2\right){}^2
   \left(q_2^2+q_{13}^2-2
   \left(q_3^2+q_{12}^2\right)\right)\right) \bigg\} \,, \nn
\eea
\bea
C^{MS}_{10}(\vec{q_1},\vec{q_2},\vec{q_3}) &=& \frac{1}{16384 \pi^2  \lambda^2(q_1,q_{13},q_3) \lambda^2(q_{12},q_{23},q_{13}) }
\bigg\{ \left(3 q_{12}^6+\left(3 q_{23}^2-5
   q_{13}^2\right) q_{12}^4+\left(q_{13}^4-18 q_{23}^2 q_{13}^2-7
   q_{23}^4\right) q_{12}^2 \right. \nn \\
&+& \left. \left(q_{13}^2-q_{23}^2\right)^2
   \left(q_{13}^2+q_{23}^2\right)\right) q_1^6+\left(-\left(7
   q_3^2+q_{13}^2\right) q_{23}^6+\left(3 q_{13}^4+\left(2 q_2^2+5
   \left(q_3^2+q_{12}^2\right)\right) q_{13}^2+41 q_3^2
   q_{12}^2\right) q_{23}^4 \right. \nn \\
&+& \left. \left(-3 q_{13}^6+\left(-4 q_2^2+11
   q_3^2+34 q_{12}^2\right) q_{13}^4+q_{12}^2 \left(12 q_2^2+34
   q_3^2-15 q_{12}^2\right) q_{13}^2-37 q_3^2 q_{12}^4\right)
   q_{23}^2 \right. \nn \\
&+&\left. \left(q_{12}^2-q_{13}^2\right)^2
   \left(q_{13}^4+\left(2 q_2^2-9 q_3^2-5 q_{12}^2\right)
   q_{13}^2+3 q_3^2 q_{12}^2\right)\right) q_1^4+\left(\left(-7
   q_3^4-18 q_{13}^2 q_3^2+q_{13}^4\right) q_{12}^6+\left(-7
   q_{13}^6 \right. \right. \nn \\
&+& \left. \left. \left(-4 q_2^2+34 q_3^2+21 q_{23}^2\right)
   q_{13}^4+q_3^2 \left(12 q_2^2+5 q_3^2+34 q_{23}^2\right)
   q_{13}^2+41 q_3^4 q_{23}^2\right) q_{12}^4+\left(11 q_{13}^8+2
   \left(4 q_2^2-7 \left(q_3^2 \right. \right. \right. \right. \nn \\
&+& \left. \left. \left. \left. q_{23}^2\right)\right)
   q_{13}^6+\left(11 q_3^4+92 q_{23}^2 q_3^2+11 q_{23}^4-24 q_2^2
   \left(q_3^2+q_{23}^2\right)\right) q_{13}^4+2 q_3^2 q_{23}^2
   \left(17 \left(q_3^2 + q_{23}^2\right)-28 q_2^2\right) q_{13}^2 \right. \right. \nn \\
&-& \left. \left. 37
   q_3^4 q_{23}^4\right)
   q_{12}^2-\left(q_{13}^2-q_{23}^2\right)^2 \left(5
   q_{13}^6+\left(4 q_2^2+2 q_3^2+q_{23}^2\right) q_{13}^4+3 q_3^2
   \left(-4 q_2^2+3 q_3^2+6 q_{23}^2\right) q_{13}^2 \right. \right. \nn \\
&-& \left. \left. 3 q_3^4
   q_{23}^2\right)\right) q_1^2+\left(q_3^2-q_{13}^2\right){}^2
   \left(\left(q_3^2+q_{13}^2\right)
   q_{12}^6+\left(q_{13}^4+\left(2 q_2^2-q_3^2-9 q_{23}^2\right)
   q_{13}^2-7 q_3^2 q_{23}^2\right) q_{12}^4-\left(5
   q_{13}^6 \right. \right. \nn \\
&+& \left. \left. \left(4 q_2^2+q_3^2+2 q_{23}^2\right) q_{13}^4+3
   q_{23}^2 \left(-4 q_2^2+6 q_3^2+3 q_{23}^2\right) q_{13}^2-3
   q_3^2 q_{23}^4\right)
   q_{12}^2+\left(q_{13}^2-q_{23}^2\right){}^2 \left(3
   q_{13}^4 \right. \right. \nn \\
&+& \left. \left. \left(2 q_2^2+q_3^2+q_{23}^2\right) q_{13}^2+3 q_3^2
   q_{23}^2\right)\right) \bigg\} \,, \nn
\eea
\bea
C^{MS}_{11}(\vec{q_1},\vec{q_2},\vec{q_3}) &=& \frac{q_{23}^2}{4096 \pi^3 \lambda^2(q_2,q_{23},q_3) \lambda^2(q_{12},q_{23},q_{13}) }
\bigg\{ \left(-q_3^4-4 q_{23}^2
   q_3^2+\left(q_2^2-q_{23}^2\right){}^2\right)
   q_{12}^6+\left(-q_3^6+\left(2 q_{23}^2 \right. \right. \nn \\
&+& \left. \left. 3
   \left(q_2^2+q_{13}^2\right)\right) q_3^4+\left(-3 q_2^4+9
   q_{23}^4-2 \left(q_2^2-4 q_{13}^2\right) q_{23}^2\right)
   q_3^2+\left(q_2^2-3 q_{13}^2-2 q_{23}^2\right)
   \left(q_2^2-q_{23}^2\right){}^2\right) q_{12}^4  \nn \\
&+&  \left(-4
   q_{23}^2 q_3^6+\left(-3 q_{13}^4+9 q_{23}^4-2 \left(q_{13}^2-4
   q_2^2\right) q_{23}^2\right) q_3^4-4 q_{23}^2 \left(q_2^4-4
   q_{23}^2 q_2^2+q_{13}^4+2 q_{23}^4 \right. \right. \nn \\
&-& \left. \left. 4 q_{13}^2
   \left(q_2^2+q_{23}^2\right)\right)
   q_3^2-\left(q_2^2-q_{23}^2\right){}^2 \left(-3 q_{13}^4+6
   q_{23}^2 q_{13}^2+q_{23}^4\right)\right)
   q_{12}^2-\left(q_{13}^2-q_{23}^2\right)^2 \left(-q_3^6+\left(3
   q_2^2 \right. \right. \nn \\
&-& \left. \left. q_{13}^2+2 q_{23}^2\right) q_3^4+\left(-3 q_2^4+6 q_{23}^2
   q_2^2+q_{23}^4\right) q_3^2+\left(q_2^2+q_{13}^2-2
   q_{23}^2\right) \left(q_2^2-q_{23}^2\right){}^2\right)+2 q_1^2
   \left(\left(q_2^4 \right. \right. \nn \\
&-&  \left. \left. 2 \left(q_3^2+q_{23}^2\right)
   q_2^2+q_3^4+q_{23}^4\right) q_{12}^4-2 \left(2 q_3^2
   q_{23}^4+q_{13}^2 \left(q_2^4-2 \left(q_3^2+q_{23}^2\right)
   q_2^2+q_3^4+q_{23}^4\right)\right)
   q_{12}^2 \right. \nn \\
&+& \left. \left(q_{13}^2-q_{23}^2\right)^2 \left(q_2^4-2
   \left(q_3^2+q_{23}^2\right)
   q_2^2+q_3^4+q_{23}^4\right)\right) \bigg\} \,.
\eea
%
%

\section{Interaction vertices}
\label{Vertices}

%
We provide here a list of the vertices used in the paper. 
The computation of the vertices  can be done by taking at most three functional derivatives of the action
with respect to the metric, since the vacuum expectation values of the fourth order derivatives correspond to massless
tadpoles, which are set to zero, as explained in the previous sections. 
We keep the notation with square brackets to indicate the flat limit of the functional derivatives in momentum space,
showing explicitly the dependence on the momenta when this occurs (which is not always the case, as for the metric tensors). We have 
\beqa
\left[\sqrt{g}\right]^{\mud\nud}
&=& 
\left[V\right]^{\mud\nud}
=
\frac{1}{2}\, \delta^{\mud\nud} \, ,
\nn \\
\left[\sqrt{g}\right]^{\mud\nud\mut\nut}
&=&
\left[V\right]^{\mud\nud\mut\nut} 
=
\frac{1}{2}\, \left(\frac{1}{2}\, \delta^{\mud\nud}\, \delta^{\mut\nut} + \left[g^{\mud\nud}\right]^{\mut\nut}\right) \, ,
\nn \\
\left[\sqrt{g}\right]^{\mud\nud\mut\nut\muq\nuq}
&=& 
\left[V\right]^{\mud\nud\mut\nut\muq\nuq}
=
\left[\sqrt{g}\right]^{\muq\nuq}\, \left[\sqrt{g}\right]^{\mud\nud\mut\nut}
+ \frac{1}{2}\, \bigg( \frac{1}{2}\, \left[g^{\mud\nud}\, g^{\mut\nut}\right]^{\muq\nuq}
+ \left[g^{\mud\nud}\right]^{\mut\nut\muq\nuq} \bigg) \, ,
\nn \\
\left[ g_{\alpha\beta} \right]^{\mud\nud}
&=& 
\frac{1}{2}\left(\delta_\alpha^{\mud} \delta_\beta^{\nud} + \delta_\alpha^{\nud} \delta_\beta^{\mud} \right)
\label{Symm} \, , 
\nn \\
\left[ g^{\alpha\beta} \right]^{\mud\nud}
&=& 
- \frac{1}{2}\left(\delta^{\alpha\mud}\delta^{\beta\nud} + \delta^{\alpha\nud}\delta^{\beta\mud}\right)  \, ,
\nn \\
\left[ g^{\alpha\beta} \right]^{\mud\nud\mut\nut}
&=& 
- \frac{1}{2}\, \bigg( \left[g^{\alpha\mud}\right]^{\mut\nut}\, \delta^{\beta\nud} 
                     + \delta^{\alpha\mud}\, \left[g^{\beta\nud}\right]^{\mut\nut}                     
                     + \left[g^{\alpha\nud}\right]^{\mut\nut}\, \delta^{\beta\mud} 
                     + \left[g^{\beta\mud}\right]^{\mut\nut}\, \delta^{\alpha\nud} \bigg) \, ,
\nn \\
\left[V_{a \lambda}\right]^{\mud\nud} 
&=& - \left[{V_a}^\lambda\right]^{\mud\nud} 
=
\frac{1}{4}\, \left({V_{a}}^{\mud}\,{\delta_\lambda}^{\nud} + {V_{a}}^{\nud}\,{\delta_\lambda}^{\mud} \right)\, , 
\nn \\
\left[V_{a \lambda}\right]^{\mud\nud\mut\nut}
&=& 
- \left[{V_a}^\lambda\right]^{\mud\nud\mut\nut} 
=
\frac{1}{4}\, \left(\left[{V_{a}}^{\mud}\right]^{\mut\nut}\,{\delta_\lambda}^{\nud} 
+ \left[{V_{a}}^{\nud}\right]^{\mut\nut}\,{\delta_\lambda}^{\mud} \right)\, , 
\nn \\
\left[V_{a \lambda}\right]^{\mud\nud\mut\nut\muq\nuq} 
&=&
-\left[{V_a}^\lambda\right]^{\mud\nud\mut\nut\mu\nuq} 
=
\frac{1}{4}\, \left(\left[{V_{a}}^{\mud}\right]^{\mut\nut\muq\nuq}\,{\delta_\lambda}^{\nud} 
+ \left[{V_{a}}^{\nud}\right]^{\mut\nut\muq\nuq}\,{\delta_\lambda}^{\mud} \right)\, .
\eeqa

For the Christoffel symbols, defined as
\beqa\label{Christoffel}
\Gamma^{\alpha}_{\beta\gamma}(z) 
&=& 
\frac{1}{2}\, g^{\alpha\kappa}(z)\, 
\left[-\pd_\kappa g_{\beta\gamma}(z) + \pd_\beta g_{\kappa\gamma}(z) + \pd_\gamma  g_{\kappa\beta}(z)\right]\, , 
\nn \\
\Gamma_{\alpha,\beta\gamma}(z) 
&=& 
\frac{1}{2}\, 
\left[-\pd_\alpha g_{\beta\gamma}(z) + \pd_\beta g_{\alpha\gamma}(z) + \pd_\gamma  g_{\alpha\beta}(z)\right]\, , 
\eeqa
we obtain
\beqa
\left[\Gamma^\alpha_{\beta\gamma}\right]^{\mud\nud}(\kd)
&=&
\frac{i}{2}\, \delta^{\alpha\lambda}\, \left(- \left[g_{\beta\gamma}\right]^{\mud\nud} k_{2\,\lambda} + 
\left[g_{\beta\lambda}\right]^{\mud\nud} k_{2\,\gamma} + \left[g_{\lambda\gamma}\right]^{\mud\nud} k_{2\,\beta} \right) \, ,
\nn \\
\left[\Gamma^\alpha_{\beta\gamma}\right]^{\mud\nud\mut\nut}(\kd,\kt)
&=&
\left[g^{\alpha\lambda}\right]^{\mud\nud}\, \left[\Gamma_{\lambda,\beta\gamma}\right]^{\mut\nut}(\kt)
\left[g^{\alpha\lambda}\right]^{\mut\nut}\, \left[\Gamma_{\lambda,\beta\gamma}\right]^{\mud\nud}(\kd) \, ,
\nn \\
\left[\Gamma^\alpha_{\beta\gamma}\right]^{\mud\nud\mut\nut\muq\nuq}(\kd,\kt,\kq)
&=&
\left[g^{\alpha\lambda}\right]^{\mud\nud}\, \left[\Gamma_{\lambda,\beta\gamma}\right]^{\mut\nut\muq\nuq}(\kt,\kq) +
\left[g^{\alpha\lambda}\right]^{\mut\nut}\, \left[\Gamma_{\lambda,\beta\gamma}\right]^{\mud\nud\muq\nuq}(\kd,\kq) \nn \\
&+&
\left[g^{\alpha\lambda}\right]^{\muq\nuq}\, \left[\Gamma_{\lambda,\beta\gamma}\right]^{\mud\nud\mut\nut}(\kd,\kt)\, .
\eeqa
It is straightforward, starting from these definitions, to derive the derivatives of the Ricci tensor
$\left[R_{\mu\nu}\right]^{\rho\sigma}(\vec{l})$ and $\left[R_{\mu\nu}\right]^{\rho\sigma\chi\omega}(\vec{l_1},\vec{l_2})$,
 defined as $R_{\mu\nu}(z) = {R^\lambda}_{\mu\lambda\nu}(z)$. We recall that in our conventions 
the Riemann tensor is defined as 
\beq
{R^\lambda}_{\mu\kappa\nu}(z) =
\pd_\nu \Gamma^\lambda_{\mu\kappa}(z) - \pd_\kappa \Gamma^\lambda_{\mu\nu}(z)
+ \Gamma^\lambda_{\nu\eta}(z)\Gamma^\eta_{\mu\kappa}(z) - \Gamma^\lambda_{\kappa\eta}(z)\Gamma^\eta_{\mu\nu}(z)\, .
\eeq

Next we list the interaction vertices for the dual theories. 
\begin{itemize}

\item \textbf{Scalar}
\beqa
V_{S\phi\phi}^{\mu\nu}(\vec{p},\vec{q})
&=&
\frac{1}{2}\left(\delta^{\mu\alpha}\delta^{\nu\beta} - \frac{1}{2}\delta^{\mu\nu}\delta^{\alpha\beta}\right)\, 
\left(p_{\alpha} q_{\beta} + p_{\beta} q_{\alpha}\right)\, ,
\nn \\
&+& 
\chi\, \left(\delta^{\mu\nu}\delta^{\alpha\beta} - \delta^{\mu\alpha}\delta^{\nu\beta}\right)\, 
\left(p_{\alpha}p_{\beta} + p_{\alpha}q_{\beta} + q_{\alpha}p_{\beta}+ q_{\alpha}q_{\beta} \right) 
\nn \\
V_{SS\phi\phi}^{\mu\nu\rho\sigma}(\vec{p},\vec{q},\vec{l})
&=&
\frac{1}{2}\, \left( \left[\sqrt{g}\right]^{\rho\sigma}\,
\left(\delta^{\mu\alpha}\delta^{\nu\beta} - \frac{1}{2}\delta^{\mu\nu}\delta^{\alpha\beta}\right)
+ \left[g^{\mu\alpha}g^{\nu\beta} - \frac{1}{2}\, g^{\mu\nu}g^{\alpha\beta} \right]^{\rho\sigma}\,\right)
\left(p_{\alpha} q_{\beta} + p_{\beta} q_{\alpha}\right)\, ,
\nn \\
&+&
\chi\,\left\{ \left( \left[\sqrt{g}\right]^{\rho\sigma}\,
\left(\delta^{\mu\nu}\delta^{\alpha\beta} - \delta^{\mu\alpha}\delta^{\nu\beta}\right)
+ \left[g^{\mu\nu}g^{\alpha\beta} - g^{\mu\alpha}g^{\nu\beta} \right]^{\rho\sigma}\,\right)
\left(p_{\alpha}p_{\beta} + p_{\alpha}q_{\beta} + p_{\beta}q_{\alpha} + q_{\alpha}q_{\beta}\right)
\right.
\nn \\
&+&
\left.
\left(\delta^{\mu\nu}\delta^{\alpha\beta} - \delta^{\mu\alpha}\delta^{\nu\beta}\right)\,
\left[\Gamma^\lambda_{\alpha\beta}\right]^{\rho\sigma}(\vec{l})\, i\, (p_\lambda + q_\lambda)
-  \left(\frac{1}{2}\, \delta^{\mu\nu}\delta^{\alpha\beta} - \delta^{\mu\alpha}\delta^{\nu\beta}\right)\, 
\left[R_{\alpha\beta}\right]^{\rho\sigma}(\vec{l}) \right\} \, ,  
\nn
\eea
\bea
V_{SSS\phi\phi}^{\mu\nu\rho\sigma\chi\omega}(\vec{p},\vec{q},\vec{l_1},\vec{l_2})
&=&
\frac{1}{2}\bigg\{\left[\sqrt{g}\right]^{\rho\sigma\chi\omega}\, 
\left(\delta^{\mu\alpha}\delta^{\nu\beta} - \frac{1}{2}\, \delta^{\mu\nu}\delta^{\alpha\beta}\right)
\nn \\
&+&
\left[\sqrt{g}\right]^{\rho\sigma}\, \left[g^{\mu\alpha}g^{\nu\beta} - \frac{1}{2}\, g^{\mu\nu} g^{\alpha\beta}\right]^{\chi\omega}	+
\left[\sqrt{g}\right]^{\chi\omega}\, \left[g^{\mu\alpha}g^{\nu\beta} - \frac{1}{2}\, g^{\mu\nu} g^{\alpha\beta}\right]^{\rho\sigma} 
\nn \\
&+&
\left[g^{\mu\alpha}g^{\nu\beta} - \frac{1}{2}\, g^{\mu\nu} g^{\alpha\beta}\right]^{\rho\sigma\chi\omega} \bigg\}\, 
\left(p_{\alpha} q_{\beta} + p_{\beta} q_{\alpha}\right)\, .
\nn \\
&+&
\chi\, \bigg\{\left[\sqrt{g}\right]^{\rho\sigma\chi\omega}\, 
\left(\delta^{\mu\nu}\delta^{\alpha\beta} - \delta^{\mu\alpha}\delta^{\nu\beta}\right)
\nn \\
&+&
\left[\sqrt{g}\right]^{\rho\sigma}\, \left[g^{\mu\nu}g^{\alpha\beta} - g^{\mu\alpha}g^{\nu\beta}\right]^{\chi\omega}	+
\left[\sqrt{g}\right]^{\chi\omega}\, \left[g^{\mu\nu}g^{\alpha\beta} - g^{\mu\alpha}g^{\nu\beta}\right]^{\rho\sigma} 
\nn \\
&+&
\left[g^{\mu\nu}g^{\alpha\beta} - g^{\mu\alpha}g^{\nu\beta}\right]^{\rho\sigma\chi\omega}
\bigg\}\, 
\left(p_\alpha p_\beta + p_\alpha q_\beta + q_\alpha p_\beta + q_\alpha q_\beta\right)
\nn \\
&+&
\chi\, \bigg\{
\left(\left[\sqrt{g}\right]^{\chi\omega}\, 
\left[\delta^{\mu\nu}\delta^{\alpha\beta} - \delta^{\mu\alpha}\delta^{\nu\beta}\right] 
+ \left[g^{\mu\nu}g^{\alpha\beta} - g^{\mu\alpha}g^{\nu\beta}\right]^{\chi\omega} \right)\, 
\left[\Gamma^\lambda_{\alpha\beta}\right]^{\rho\sigma}(\vec{l_1}) 
\nn \\
&+&
\left(\rho,\sigma,l_1\right) \leftrightarrow \left(\tau,\omega,l_2\right) 
+ \left(\delta^{\mu\nu}\delta^{\alpha\beta} - \delta^{\mu\alpha}\delta^{\nu\beta}\right)\, 
\left[\Gamma^\lambda_{\alpha\beta}\right]^{\rho\sigma\chi\omega}(\vec{l_1},\vec{l_2})
\bigg\}\, i\, \left(p_\lambda + q_\lambda\right)
\nn \\
&+&
\chi \bigg\{ 
\left(\left[\sqrt{g}\right]^{\chi\omega}\, \left( \delta^{\mu\alpha}\delta^{\nu\beta} -
\frac{1}{2}\, \delta^{\mu\nu}\delta^{\alpha\beta}\right)+
\left[g^{\mu\alpha}g^{\nu\beta} - \frac{1}{2}\,g^{\mu\nu}g^{\alpha\beta}\right]^{\chi\omega} \right)
\, \left[R_{\alpha\beta}\right]^{\rho\sigma}(\vec{l_1})
\nn \\
&+& 
\left(\rho,\sigma,l_1\right) \leftrightarrow \left(\tau,\omega,l_2\right) 
+\left(\delta^{\mu\alpha}\delta^{\nu\beta} - \frac{1}{2}\,\delta^{\mu\nu}\delta^{\alpha\beta}\right) 
\left[R_{\alpha\beta}\right]^{\rho\sigma\chi\omega}(\vec{l_1},\vec{l_2}) \bigg\}
\eeqa
\\ \\
\item \textbf{Fermion} 

\beqa
V^{\mu\nu}_{S\bar\psi\psi}(\vec{p},\vec{q}) 
&=&
\frac{1}{2}\, \left( \left[V\right]^{\mu\nu}\, {\delta_a}^{\lambda} + \left[{V_a}^\lambda\right]^{\mu\nu} \right)\, 
\gamma^a\, (p_\lambda - q_\lambda)\, ,
\nn \\
V^{\mu\nu\rho\sigma}_{SS\bar\psi\psi}(\vec{p},\vec{q},\vec{l_1})
&=&
\frac{1}{2}\, \bigg( \left[V\right]^{\mu\nu\rho\sigma}\, {\delta_a}^{\lambda}
+ \left[V\right]^{\mu\nu}\, \left[{V_a}^{\lambda}\right]^{\rho\sigma} 
+ \left[V\right]^{\rho\sigma}\, \left[{V_a}^{\lambda}\right]^{\mu\nu}
+ \left[{V_a}^{\lambda}\right]^{\mu\nu\rho\sigma}\bigg)\, \gamma^a\, (p_\lambda - q_\lambda)
\nn \\
&+&
\frac{1}{16}\, \bigg( \left[V\right]^{\mu\nu}\, {\delta_a}^\lambda + \left[{V_a}^\lambda\right]^{\mu\nu}\bigg)\,
\left\{\gamma^a,\left[\gamma^b,\gamma^c\right]\right\}\, \left[\Omega_{bc,\lambda}\right]^{\rho\sigma}(\vec{l_1})\, ,
\nn
\eeqa
\beqa
V^{\mu\nu\rho\sigma\chi\omega}_{SSS\bar\psi\psi}(\vec{p},\vec{q},\vec{l_1},\vec{l_2})
&=&
\frac{1}{2}\, \bigg( 
  \left[V\right]^{\mu\nu\rho\sigma\chi\omega}\, {\delta_a}^{\lambda}
+ \left[V\right]^{\mu\nu\rho\sigma}\, \left[{V_a}^{\lambda}\right]^{\chi\omega}
+ \left[V\right]^{\mu\nu\chi\omega}\, \left[{V_a}^{\lambda}\right]^{\rho\sigma} \nn \\
&+&
  \left[V\right]^{\rho\sigma\chi\omega}\, \left[{V_a}^{\lambda}\right]^{\mu\nu}
+ \left[V\right]^{\mu\nu}\, \left[{V_a}^{\lambda}\right]^{\rho\sigma\chi\omega}  
+ \left[V\right]^{\rho\sigma}\, \left[{V_a}^{\lambda}\right]^{\mu\nu\chi\omega} \nn \\
&+&
  \left[V\right]^{\chi\omega}\, \left[{V_a}^{\lambda}\right]^{\mu\nu\rho\sigma}
+ \left[{V_a}^{\lambda}\right]^{\mu\nu\rho\sigma\chi\omega}\bigg)\, \gamma^a\, (p_\lambda - q_\lambda) 
\nn \\
&+&
\frac{1}{16}\,\{\gamma^a,[\gamma^b,\gamma^c]\}\,
\bigg\{ \bigg(
\left[V\right]^{\mu\nu\rho\sigma}\, {\delta_a}^{\lambda}
+ \left[V\right]^{\mu\nu}\, \left[{V_a}^{\lambda}\right]^{\rho\sigma}
\nn \\
&+&
\left[V\right]^{\rho\sigma}\, \left[{V_a}^{\lambda}\right]^{\mu\nu}
+ \left[{V_a}^{\lambda}\right]^{\mu\nu\rho\sigma} \bigg)
\left[\Omega_{bc,\lambda}\right]^{\chi\omega}(\vec{l_2})
\nn \\
&+&
\bigg( \left[V\right]^{\mu\nu}\, {\delta_a}^\lambda + \left[{V_a}^\lambda\right]^{\mu\nu}\bigg)\,
\left[\Omega_{bc,\lambda}\right]^{\rho\sigma\chi\omega}(\vec{l_1},\vec{l_2})\bigg\}\, ,
\eeqa
where the spin connection was defined in (\ref{SpinCon}).
%
\item \textbf{Gauge field} 

%
\beqa
V^{\mu\nu\,\tau\vartheta}_{SAA}(\vec{p},\vec{q})
&=&
\frac{1}{2}\, \left(\delta^{\mu\lambda}\delta^{\alpha\kappa}\delta^{\nu\beta} 
+ \frac{1}{4}\delta^{\mu\nu}\delta^{\alpha\lambda}\delta^{\beta\kappa}\right)\,
\left[F_{\alpha\beta}F_{\lambda\kappa}\right]^{\tau\vartheta}(\vec{p},\vec{q})
\nn \\
V^{\mu\nu\rho\sigma\,\tau\vartheta}_{SSAA}(\vec{p},\vec{q})
&=&
\frac{1}{2}\, \left\{ \left[\sqrt{g}\right]^{\rho\sigma}\, 
\left(\delta^{\mu\lambda}\delta^{\alpha\kappa}\delta^{\nu\beta} 
+ \frac{1}{4}\delta^{\mu\nu}\delta^{\alpha\lambda}\delta^{\beta\kappa}\right)
\right.
\nn \\
&+&
\left.
\left[ g^{\mu\rho}g^{\alpha\sigma}g^{\nu\beta} + \frac{1}{4}\, g^{\mu\nu}g^{\alpha\beta}g^{\rho\sigma}\right]^{\rho\sigma}
\right\}\, \left[F_{\alpha\beta}F_{\lambda\kappa}\right]^{\tau\vartheta}(\vec{p},\vec{q})
\nn\\
V^{\mu\nu\rho\sigma\chi\omega\, \tau\vartheta}_{SSSAA}(\vec{p},\vec{q})
&=&
\frac{1}{2}\, \left\{
\left[\sqrt{g}\right]^{\rho\sigma\chi\omega}\,
\left(\delta^{\mu\lambda}\delta^{\alpha\kappa}\delta^{\nu\beta} 
+ \frac{1}{4}\delta^{\mu\nu}\delta^{\alpha\lambda}\delta^{\beta\kappa}\right)
\right.
\nn \\
&+&
\left.
\left[\sqrt{g}\right]^{\rho\sigma}\, 
\left[g^{\mu\lambda}g^{\alpha\kappa}g^{\nu\beta} + \frac{1}{4}g^{\mu\nu}g^{\alpha\lambda}g^{\beta\kappa}\right]^{\chi\omega} +
\left[\sqrt{g}\right]^{\chi\omega}\, 
\left[g^{\mu\lambda}g^{\alpha\kappa}g^{\nu\beta} + \frac{1}{4}g^{\mu\nu}g^{\alpha\lambda}g^{\beta\kappa}\right]^{\rho\sigma}
\right.
\nn \\
&+& 
\left.
\left[ g^{\mu\lambda}g^{\alpha\kappa}g^{\nu\beta} 
+ \frac{1}{4}g^{\mu\nu}g^{\alpha\lambda}g^{\beta\kappa}\right]^{\rho\sigma\chi\omega}
\right\}\, \left[F_{\alpha\beta}F_{\lambda\kappa}\right]^{\tau\vartheta}(\vec{p},\vec{q})
\eeqa
where we have introduced
\beqa
\left[F_{\alpha\beta}F_{\lambda\kappa}\right]^{\tau\vartheta}(\vec{p},\vec{q})
&=& 
\int\, d^4x d^4y\, e^{- i\, p\cdot x -i\, q\cdot y}\, 
\frac{\delta^2 \left(F_{\alpha\beta}(0)F_{\lambda\kappa}(0)\right)}{\delta A_{\tau}(x)\delta A_{\vartheta}(y)}
\nn \\
&=&
-  \left(
\delta^{\tau}_{\lambda}\, \delta^{\vartheta}_{\alpha}\, p_{\kappa}\, q_{\beta}
- \delta^{\tau}_{\lambda}\, \delta^{\vartheta}_{\beta}\, p_{\kappa}\, q_{\alpha}
- \delta^{\tau}_{\kappa}\, \delta^{\vartheta}_{\alpha} p_{\lambda}\, q_{\beta}
+ \delta^{\tau}_{\kappa}\, \delta^{\vartheta}_{\beta}\, p_{\lambda}\, q_{\alpha}
+  \left(\tau,\vec{p}\right) \leftrightarrow  \left(\vartheta,\vec{q}\right)
\right) \,.
\eeqa

\end{itemize}

\end{appendix}


\end{document}